\documentclass[zpreprint,zbstdefault]{zeus_paper}
\usepackage[english]{babel}
\newcommand{\ZcoosysA}{%
The ZEUS coordinate system is a right-handed Cartesian system, with the $Z$
axis pointing in the proton beam direction, referred to as the ``forward
direction'', and the $X$ axis pointing left towards the center of HERA.
The coordinate origin is at the nominal interaction point.\xspace}

\newcommand{\ZcoosysfnA}{\footnote{\ZcoosysA}}

\newcommand{\Zdetdesc}{%
A detailed description of the ZEUS detector can be found 
elsewhere~\cite{zeus:1993:bluebook}. A brief outline of the 
components that are most relevant for this analysis is given
below.\xspace}
\newcommand{\Zctddesc}[1]{%
Charged particles are tracked in the central tracking detector (CTD)~\citeCTD,
which operates in a magnetic field of $1.43\Tesla$ provided by a thin 
superconducting coil. The CTD consists of 72~cylindrical drift chamber 
layers, organized in 9~superlayers covering the polar-angle#1 region 
\mbox{$15^\circ<\theta<164^\circ$}. The transverse-momentum resolution for
full-length tracks is $\sigma(p_T)/p_T=0.0058p_T\oplus0.0065\oplus0.0014/p_T$,
with $p_T$ in $\Gev$.}
\newcommand{\Zcaldesc}{%
The high-resolution uranium--scintillator calorimeter (CAL)~\citeCAL consists 
of three parts: the forward (FCAL), the barrel (BCAL) and the rear (RCAL)
calorimeters. Each part is subdivided transversely into towers and
longitudinally into one electromagnetic section and either one (in RCAL)
or two (in BCAL and FCAL) hadronic sections. The smallest subdivision of
the calorimeter is called a cell.  The CAL energy resolutions, as measured under
test-beam conditions, are $\sigma(E)/E=0.18/\sqrt{E}$ for electrons and
$\sigma(E)/E=0.35/\sqrt{E}$ for hadrons, with $E$ in $\Gev$.}

\chardef\usc=95
\chardef\til=126
\catcode`\@=11 
\DeclareRobustCommand\xdotspace{\futurelet\@let@token\@xdotspace}
\def\@xdotspace{%
  \ifx\@let@token.\else
  \ifx\@let@token\bgroup.\else
  \ifx\@let@token\egroup.\else
  \ifx\@let@token\/.\else
  \ifx\@let@token\ .\else
  \ifx\@let@token~.\else
  \ifx\@let@token!.\else
  \ifx\@let@token,.\else
  \ifx\@let@token:.\else
  \ifx\@let@token;.\else
  \ifx\@let@token?.\else
  \ifx\@let@token/.\else
  \ifx\@let@token'.\else
  \ifx\@let@token).\else
  \ifx\@let@token-.\else
  \ifx\@let@token\@xobeysp.\else
  \ifx\@let@token\space.\else
  \ifx\@let@token\@sptoken.\else
   .\space
   \fi\fi\fi\fi\fi\fi\fi\fi\fi\fi\fi\fi\fi\fi\fi\fi\fi\fi}
\catcode`\@=12 

\newcommand{\stru}[2]{%
   \relax\ifmmode\hbox{\vrule height#1 depth#2 width0pt}%
   \else\vrule height#1 depth#2 width0pt\fi}

\newcommand{\Ronum}[1]{\uppercase\expandafter{\romannumeral#1}}
\newcommand{\ronum}[1]{\expandafter{\romannumeral#1}}
\DeclareRobustCommand{\LaTeXZ}{%
  \LaTeX\kern-.05em4\kern-.1em
  {\raisebox{-0.2ex}{$\scriptstyle\text{ZEUS}$}}\xspace}

\DeclareMathAlphabet{\mathbf}{OT1}{cmr}{bx}{sl}
\newcommand{\eVdist}{\kern-0.06667em}

\newcommand{\Gev}{{\text{Ge}\eVdist\text{V\/}}}

\newcommand{\mev}{{\,\text{Me}\eVdist\text{V\/}}}
\newcommand{\gev}{{\,\text{Ge}\eVdist\text{V\/}}}

\newcommand{\pb}{\,\text{pb}}

\newcommand{\cm}{\,\text{cm}}

\newcommand{\Tesla}{\,\text{T}}

\newcommand{\slashfrac}[2]{%
  \raisebox{0.5ex}{\ensuremath #1}\kern-0.12em/\kern-0.08em
  \raisebox{-.8ex}{\ensuremath #2}}

\newcommand{\sqr}[3]{%
    {\vcenter{\hrule height.#3ex\hbox{\vrule width.#2ex height#1ex
     \kern#1ex\vrule width.#3ex}\hrule height.#2ex}}}

\catcode`\@=11 
\newcommand{\parenbar}{\mathpalette\p@renb@r}
\def\p@renb@r#1#2{\vbox{%
  \ifx#1\scriptscriptstyle \dimen@.7em\dimen@ii.2em\else
  \ifx#1\scriptstyle \dimen@.8em\dimen@ii.25em\else
  \dimen@1em\dimen@ii.4em\fi\fi \offinterlineskip
  \ialign{\hfill##\hfill\cr
    \vbox{\hrule width\dimen@ii}\cr
    \noalign{\vskip-.3ex}%
    \hbox to\dimen@{$\mathchar300\hfil\mathchar301$}\cr
    \noalign{\vskip-.3ex}%
    $#1#2$\cr}}}
\catcode`\@=12 

\newcommand{\rnge}{\hbox{$\,\text{--}\,$}}

\newcommand{\IP}{{\rm I$\kern-0.01667em$P}\xspace}

\mathchardef\qsm=63
\mathchardef\pls=43
\mathchardef\mns=512
\mathchardef\plm=518
\mathchardef\eql=61
\mathchardef\smallleft=300
\mathchardef\smallright=301
\mathchardef\les=316
\mathchardef\gre=318
\mathchardef\leq=532
\mathchardef\grq=533

\catcode`\@=11 
\newcounter{pict@width}
\newcounter{pict@height}
\newlength{\pict@scale}
\newcommand{\psfigadd}[4]{%
\setcounter{pict@width}{1*\ratio{#2+\pict@scale/2}{\pict@scale}}
\setcounter{pict@height}{1*\ratio{#3+\pict@scale/2}{\pict@scale}}
\setlength{\unitlength}{\pict@scale}
\hbox to #2{\hspace{-\fill}\begin{picture}(\thepict@width,\thepict@height)
\put(0,0){\psfig{figure=#1,width=#2,height=#3,clip=}}
\SetScale{0.283466457}
\SetWidth{1.763889}
{#4}
\end{picture}}
}
\newcounter{pict@widthfst}
\newcounter{pict@widthscd}
\newcounter{pict@widthtot}
\newcommand{\psfigaddtwo}[7]{%
\setcounter{pict@widthfst}{1*\ratio{#2+\pict@scale/2}{\pict@scale}}
\setcounter{pict@widthscd}{1*\ratio{#2+#4+\pict@scale/2}{\pict@scale}}
\setcounter{pict@widthtot}{1*\ratio{#2+#4+#6+\pict@scale/2}{\pict@scale}}
\setcounter{pict@height}{1*\ratio{#3+\pict@scale/2}{\pict@scale}}
\setlength{\unitlength}{\pict@scale}
\hbox{\hspace{-\fill}\begin{picture}(\thepict@widthtot,\thepict@height)
\put(0,0){\psfig{figure=#1,width=#2,height=#3,clip=}}
\put(\thepict@widthscd,0){\psfig{figure=#5,width=#6,height=#3,clip=}}
\SetScale{0.283466457}
\SetWidth{1.763889}
{#7}
\end{picture}}
}
\newcommand{\psfigror}[4]{%
\setcounter{pict@width}{1*\ratio{#2+\pict@scale/2}{\pict@scale}}
\setcounter{pict@height}{1*\ratio{#3+\pict@scale/2}{\pict@scale}}
\setlength{\unitlength}{\pict@scale}
\hbox{\begin{picture}(\thepict@width,\thepict@height)
\put(0,\thepict@height){\psfig{figure=#1,width=#3,height=#2,clip=,angle=270}}
\SetScale{0.283466457}
\SetWidth{1.763889}
{#4}
\end{picture}}
}
\newcommand{\psfigrol}[4]{%
\setcounter{pict@width}{1*\ratio{#2+\pict@scale/2}{\pict@scale}}
\setcounter{pict@height}{1*\ratio{#3+\pict@scale/2}{\pict@scale}}
\setlength{\unitlength}{\pict@scale}
\hbox{\begin{picture}(\thepict@width,\thepict@height)
\put(0,0){\psfig{figure=#1,width=#3,height=#2,clip=,angle=90}}
\SetScale{0.283466457}
\SetWidth{1.763889}
{#4}
\end{picture}}
}
\catcode`\@=12 
\newlength\listtextwidth

\catcode`\@=11 
\newlength{\@tabfninsert}
\newlength{\@tabfnwidth}
\newcommand{\tabfootnote}[2]{%
  \setlength{\@tabfninsert}{0.8em}
  \setlength{\@tabfnwidth}{\textwidth}
  \addtolength{\@tabfnwidth}{-\@tabfninsert}
  \addtolength{\@tabfnwidth}{-0.4em}
  \noindent\makebox[\@tabfninsert][r]{\footnotesize$^{#1}$\hfil}\hfill%
  \parbox[t]{\@tabfnwidth}{\footnotesize #2\hfill}}
\catcode`\@=12 

\def\citeCTD{{\cite{%
nim:a279:290,*npps:b32:181,*nim:a338:254%
}}\xspace}
\def\citeCAL{{\cite{%
nim:a309:77,*nim:a309:101,*nim:a321:356,*nim:a336:23%
}}\xspace}

\begin{document}
\prepnum{DESY--06--226}

\title{
Measurement of $K^{0}_{S}$, $\Lambda$ and $\bar{\Lambda}$ production \\ at HERA
}                                                       

\author{ZEUS Collaboration}
\draftversion{DESY--06--226}
\date{December 2006}


\abstract{
The production of the neutral strange hadrons $K^{0}_{S}$, $\Lambda$ and $\bar{\Lambda}$ has been measured 
in $ep$ collisions at HERA using the ZEUS detector.  
Cross sections, baryon-to-meson ratios,
relative yields of strange and charged light hadrons, $\Lambda$ ($\bar{\Lambda}$) 
asymmetry and polarization have been measured in three kinematic regions: 
$Q^2 > 25 \gev^2$; $5 < Q^2 < 25 \gev^2$; and in photoproduction ($Q^2 \simeq 0$). In photoproduction 
the presence of two hadronic jets, each with at least $5 \gev$ transverse energy, was required.
The measurements agree in general with Monte Carlo models and 
are consistent with measurements made at $e^+ e^-$
colliders, except for an enhancement of baryon relative to meson production in 
photoproduction.
}

\makezeustitle

\def\3{\ss}                                                                                        
\pagenumbering{Roman}                                                                              
                                                   %
\begin{center}                                                                                     
{                      \Large  The ZEUS Collaboration              }                               
\end{center}                                                                                       
  S.~Chekanov$^{   1}$,                                                                            
  M.~Derrick,                                                                                      
  S.~Magill,                                                                                       
  S.~Miglioranzi$^{   2}$,                                                                         
  B.~Musgrave,                                                                                     
  D.~Nicholass$^{   2}$,                                                                           
  \mbox{J.~Repond},                                                                                
  R.~Yoshida\\                                                                                     
 {\it Argonne National Laboratory, Argonne, Illinois 60439-4815}, USA~$^{n}$                       
\par \filbreak                                                                                     
  M.C.K.~Mattingly \\                                                                              
 {\it Andrews University, Berrien Springs, Michigan 49104-0380}, USA                               
\par \filbreak                                                                                     
  M.~Jechow, N.~Pavel~$^{\dagger}$, A.G.~Yag\"ues Molina \\                                        
  {\it Institut f\"ur Physik der Humboldt-Universit\"at zu Berlin,                                 
           Berlin, Germany}
% ...
\newpage                                                                                           
$^{\    1}$ supported by DESY, Germany \\                                                          
$^{\    2}$ also affiliated with University College London, UK \\                                  
% ...
$^{\dagger}$ deceased \\                                                                           
%
\newpage   
                                                           %
                                                           %
\begin{tabular}[h]{rp{14cm}}                                                                       
$^{a}$ &  supported by the Natural Sciences and Engineering Research Council of Canada (NSERC) \\  
$^{b}$ &  supported by the German Federal Ministry for Education and Research (BMBF), under        
          contract numbers HZ1GUA 2, HZ1GUB 0, HZ1PDA 5, HZ1VFA 5\\                                
$^{c}$ &  supported in part by the MINERVA Gesellschaft f\"ur Forschung GmbH, the Israel Science   
          Foundation (grant no. 293/02-11.2) and the U.S.-Israel Binational Science Foundation \\  
$^{d}$ &  supported by the German-Israeli Foundation and the Israel Science Foundation\\           
$^{e}$ &  supported by the Italian National Institute for Nuclear Physics (INFN) \\                
$^{f}$ &  supported by the Japanese Ministry of Education, Culture, Sports, Science and Technology 
          (MEXT) and its grants for Scientific Research\\                                          
$^{g}$ &  supported by the Korean Ministry of Education and Korea Science and Engineering          
          Foundation\\                                                                             
$^{h}$ &  supported by the Netherlands Foundation for Research on Matter (FOM)\\                   
$^{i}$ &  supported by the Polish State Committee for Scientific Research, grant no.               
          620/E-77/SPB/DESY/P-03/DZ 117/2003-2005 and grant no. 1P03B07427/2004-2006\\             
$^{j}$ &  partially supported by the German Federal Ministry for Education and Research (BMBF)\\   
$^{k}$ &  supported by RF Presidential grant N 1685.2003.2 for the leading scientific schools and  
          by the Russian Ministry of Education and Science through its grant for Scientific        
          Research on High Energy Physics\\                                                        
$^{l}$ &  supported by the Spanish Ministry of Education and Science through funds provided by     
          CICYT\\                                                                                  
$^{m}$ &  supported by the Particle Physics and Astronomy Research Council, UK\\                   
$^{n}$ &  supported by the US Department of Energy\\                                               
$^{o}$ &  supported by the US National Science Foundation. Any opinion,                            
findings and conclusions or recommendations expressed in this material                             
are those of the authors and do not necessarily reflect the views of the                           
National Science Foundation.\\                                                                     
$^{p}$ &  supported by the Polish Ministry of Science and Higher Education\\                       
$^{q}$ &  supported by FNRS and its associated funds (IISN and FRIA) and by an Inter-University    
          Attraction Poles Programme subsidised by the Belgian Federal Science Policy Office\\     
$^{r}$ &  supported by the Malaysian Ministry of Science, Technology and                           
Innovation/Akademi Sains Malaysia grant SAGA 66-02-03-0048\\                                       
\end{tabular}                                                                                      
                                                           %
                                                           %

\pagenumbering{arabic}
\pagestyle{plain}
\section{Introduction}
\label{sec-int}

Production of $K^{0}_{S}$, $\Lambda$ and $\bar{\Lambda}$ hadrons has been extensively studied at particle colliders: $e^+ e^-$ \cite{zfp:c27:27,*zfp:c47:167,*prl:54:274,*prl:54:2071,*pr:d31:3013,*pr:d35:2639,*pl:b105:75,*zfp:c45:209,*zfp:c46:397,*zfp:c64:361,*pl:b318:249,zfp:c65:587,pl:b328:223,*pl:b291:503,*zfp:c67:389}, $ep$ \cite{zfp:c68:29,*epj:c2:77,*np:b480:3,*zfp:c76:213}, $p \bar{p}$ \cite{zfp:c41:179,*np:b328:36,*pr:d72:052001,pl:b366:441} and $pp$ \cite{nucl-ex-0607033}. The data have been used to test QCD and build phenomenological models extending QCD predictions beyond what can be calculated from first principles.  

The results on $K^{0}_{S}$, $\Lambda$, and $\bar{\Lambda}$ production presented in this paper are based on a data sample of 121$\pb^{-1}$ collected by the ZEUS experiment at HERA, about 100 times larger than used in previous HERA publications \cite{zfp:c68:29,*epj:c2:77,*np:b480:3,*zfp:c76:213} and extend the kinematical region of the measurements, thereby providing a tighter constraint on models. 

The measurements have been performed in three different regions of $Q^{2}$, where $Q^{2}$ is the virtuality of 
the exchanged boson:  Deep Inelastic Scattering (DIS) with $Q^2 > 25 \gev^2$; DIS with $5 < Q^2 < 25 \gev^2$; 
and photoproduction, $Q^2 \simeq 0 \gev^2$, in which a quasi-real photon interacts with the proton. In the photoproduction 
sample, two jets, each of at least $5 \gev$ transverse energy, were required.

The following measurements are presented in this paper: differential cross sections, baryon-antibaryon asymmetry, 
baryon-to-meson ratio, ratio of strange-to-light hadrons, and the $\Lambda$ and $\bar{\Lambda}$ transverse spin 
polarization. There was no attempt to separate direct production from resonance decays: all sources for $K^{0}_{S}$, 
$\Lambda$, and $\bar{\Lambda}$ production were included. These measurements are relevant for modeling production 
of hadrons at high energies, for example in Monte Carlo (MC) programs 
\cite{cpc:71:15, cpc:101:108, cpc:86:147, cpc:135:238, hep-ph-0210213}, and for testing the mechanism 
for baryon transport along the rapidity axis \cite{zfp:c75:693}, the mechanisms for baryon 
production \cite{prl:90:202303,*prl:91:052302}, effects due to QCD instantons 
\cite{pl:b59:85,prl:37:8,*pr:d14:3432,*pr:d18:2199,*prep:142:357,pl:b188:506,*np:b330:1,*np:b343:310, 
*np:b365:3,*zfp:c66:285,*pl:b555:227} and the mechanisms for the transverse spin polarization of hadrons 
\cite{pl:b105:403,ijmp:a5:1197,pl:b183:357}.

\section{Experimental setup}
The data were collected by the ZEUS detector at the HERA $ep$ collider during the running period 1996\rnge2000. 
The data correspond to an integrated luminosity of 121$\pb^{-1}$, of which 82$\pb^{-1}$ 
were collected at $\sqrt{s} = 318 \gev$ (the electron or 
positron\footnote{In the text, electron beam, as well as scattered electron, apply to both electron and positron.} 
beam energy, $E^\mathrm{beam}_\mathrm{e}$, was 27.5 GeV and the proton beam 
energy was 920 GeV) and 39$\pb^{-1}$ at $\sqrt{s} = 300 \gev$ (where the proton beam energy was 820 GeV).

\Zdetdesc 

\Zctddesc{\ZcoosysfnA}

\Zcaldesc

A three-level trigger system was used to select events on-line \cite{zeus:1993:bluebook}. 
At the third level, DIS events were accepted on the basis of the identification of a scattered 
electron candidate using localised energy deposition in the CAL. 
As there was no possibility to select inclusive photoproduction sample, 
the requirement for photoproduction events \cite{epj:c1:109} was based on running 
a jet 
algorithm  
using the energies and positions of the CAL cells. 
Events with at least two jets were accepted, where each jet was required 
to have transverse energy greater than 4.5 $\gev$ and 
pseudorapidity \footnote{The pseudorapidity $\eta$ is defined as $\eta \equiv - \ln \tan(\theta/2)$, 
where $\theta$ is a scattering angle.}
less than 2.5.

The luminosity was measured using the bremsstrahlung process $ep \rightarrow ep\gamma$ with the luminosity monitor \cite{desy-92-066,*zfp:c63:391,*acpp:b32:2025}, a lead-scintillator calorimeter placed in the HERA tunnel at $Z = -107$ m.

\section{Event reconstruction and selection}
\subsection{Deep inelastic scattering sample}
The DIS events are characterised by a scattered electron detected in the CAL. The scattered electron was identified 
from the energy deposit in the CAL using a neural network \cite{nim:a365:508,*nim:a391:360}. The Bjorken variable 
$x_\mathrm{Bj}$ \cite{pr:185:1969} and $Q^2$ were reconstructed using the double angle method (DA) 
\cite{proc:hera:1991:23,*proc:hera:1991:43} which has the best resolution in the $Q^2$ region studied. 
The inelasticity variable, $y$, was reconstructed using both the electron (e) \cite{proc:hera:1991:23,*proc:hera:1991:43} 
and Jacquet-Blondel (JB) \cite{proc:epfacility:1979:391} methods. The following requirements were applied 
offline to select events with $Q^{2} > 25 \gev^2$ (called the high-$Q^{2}$ sample):
\begin{itemize}
\item $\mid Z_{\mathrm{vtx}}\mid < 50 \cm$ to reduce the background from non-$ep$ collisions;
\item $38 < \delta < 65 \gev$, where $\delta = \sum_i(E_i-P_{Z,i})$ and the sum runs over the energy and 
longitudinal momentum of all CAL cell deposits. This cut reduced the background from photoproduction and events with 
large radiative corrections;
\item an identified scattered electron with energy above 10\gev;
\item the impact position of the scattered electron on the CAL satisfied $\sqrt{X^2+Y^2} > 36 \cm$;
\item the electron was isolated: the energy from all CAL cell deposits not associated with 
the scattered electron but in an $\eta - \phi$ cone of radius 0.8 centered on the electron was required 
to be below 5\gev. This requirement reduced photoproduction background;
\item a track match with any electron falling in the range $0.3 < \theta < 2.6$, 
well within the CTD acceptance. For $\theta$ outside this region, $\delta > 44 \gev$ was required. 
This cut further suppressed events from non-$ep$ interaction and photoproduction;
\item $y_\mathrm{JB}>0.02$ to improve the accuracy of the DA reconstruction;
\item $y_\mathrm{e}<0.95$ to remove events where fake electrons were found in the FCAL;
\item $Q^2_\mathrm{DA} > 25 \gev^2$.
\end{itemize}
  
The same selection was used to obtain the low-$Q^{2}$ DIS sample, except for the $Q^{2}$ requirement and 
the position of the scattered electron, which were as follows:
\begin{itemize}
\item the impact position of the scattered electron on the CAL was required to be
outside a rectangle of dimensions $26 \times 14 \cm^2$, 
centred on the beam pipe;
\item $5 < Q^{2}_\mathrm{DA} < 25 \gev^2$.
\end{itemize}
The trigger for selecting low-$Q^{2}$ events was normally prescaled, so the data correspond to an 
integrated luminosity of 16.6 $\pb^{-1}$.  

It should be noted that there was no jet requirement in the DIS event samples.
\subsection{Photoproduction sample}
Photoproduction events were selected applying the following criteria, described in an earlier publication \cite{epj:c1:109}:
\begin{itemize}
\item $|Z_\mathrm{vtx}| < 50 \cm$, to reduce background from non-$ep$ collisions;
\item $y_\mathrm{JB} > 0.2 $, to further reduce background from non-$ep$ collisions;
\item events were removed where an electron was found with $y_\mathrm{e} < 0.85$, 
reducing the background from neutral current DIS events;
\item $y_\mathrm{JB} < 0.85 $, to reduce background from neutral current DIS events where the electron was not identified;
\item charged current DIS events were removed by rejecting events with 
$ P_{T}^\mathrm{miss} / \sqrt{E_{T}} > 2.0 \gev^{\frac{1}{2}}$, where $P_{T}^\mathrm{miss}$ is the missing transverse momentum and $E_T$ is the total transverse energy.
\end{itemize}

Energy Flow Objects, reconstructed from a combination of calorimeter and tracking information to give the best resolution of kinematic variables, were used as the input to the $k_{T}$ cluster jet-finding algorithm \cite{pl:b285:291,*np:b406:187}, which was run in the 
longitudinally invariant inclusive mode \cite{pr:d48:3160}. The transverse energy of the jets was corrected 
for energy losses in inactive material in front of the CAL, as described in a previous publication \cite{epj:c11:35}. 
An event was accepted if it contained at least two jets, both satisfying the following criteria:
\begin{itemize}
\item jet transverse energy $E^{\mathrm{jet}}_{T} > 5 \gev$;
\item jet pseudorapidity $|\eta^{\mathrm{jet}}| < 2.4$.
\end{itemize}
  


Photoproduction events selected in this way contributed about 10\% to the total photoproduction cross section. 
The photoproduction sample was divided into subsamples
using
the variable $x_{\gamma}^\mathrm{OBS}$, 
a measure of the fraction of the photon energy transferred to the dijet system,
defined as:
\begin{equation*} \label{x_gamma_obs}
x_{\gamma}^\mathrm{OBS} = \frac{\sum E_{T}^{\mathrm{jet}} e^{-\eta^{\mathrm{jet}}}}{2y_\mathrm{JB}E^\mathrm{beam}_\mathrm{e}}
\end{equation*}
where the sum runs over the two jets with highest transverse energy. In leading-order QCD, $x_{\gamma}^\mathrm{OBS} = 1$ corresponds to direct photon processes in which the photon takes part in the hard scattering as a point-like particle. Resolved photon processes, in which the photon acts as a source of partons, populate the region at $x_{\gamma}^\mathrm{OBS} < 1$. The sample with $x_{\gamma}^\mathrm{OBS} > 0.75$ is classified as direct-enriched, and that with $x_{\gamma}^\mathrm{OBS} < 0.75$ as resolved-enriched.

\subsection{Strange particle reconstruction}

Candidates for long-lived neutral strange hadrons decaying to two charged particles were identified 
by selecting pairs of oppositely charged tracks, fitted to a displaced secondary vertex. Events were 
required to have at least one such candidate. These secondary vertices were found by the ZEUS 
track-finding software, which is based on minimizing the $\chi^2$ arising from fitting tracks to vertices 
\cite{nim:a311:139}. Displaced vertices were typically more than $3 \cm$ away from the primary vertex. 
The minimal distance required to resolve a displaced vertex from the primary vertex was about 1 cm. 
The tracks fitted to this vertex were required to pass 
through at least the fifth superlayer of the CTD, the transverse momentum was required to be greater 
than $150 \mev$ and the absolute pseudorapidity in the laboratory frame was required to be less than 1.5. 
These constraints ensured a good track resolution and acceptance. The $\Lambda$, $\bar{\Lambda}$ 
and $K^{0}_{S}$ particles may also be created in interactions with the beam pipe. To remove these events 
a collinearity cut on the angle between the reconstructed  candidate momentum and the vector 
joining the primary vertex to secondary vertex was applied. This angle was restricted to be less than 0.2.

The $\Lambda$($\bar{\Lambda}$) candidates were reconstructed by their charged decay mode to $p\pi^{-}$( $\bar{p}\pi^{+}$)
(branching ratio 63.9 $\pm$ 0.5\% \cite{pl:b592:1}). The track with the larger momentum was assigned the mass 
of the proton, while the other was assigned the mass of the charged pion, as the decay proton always has 
a larger momentum than the pion, provided the  $\Lambda$($\bar{\Lambda}$) momentum is greater than $0.3 \gev$. 
Additional requirements to select $\Lambda(\bar{\Lambda})$ are given in the following:
\begin{itemize}
\item $0.6 < P^\mathrm{LAB}_T(\Lambda,\,\bar{\Lambda}) < 2.5\gev$, where $P^\mathrm{LAB}_T(\Lambda,\,\bar{\Lambda})$ 
is the transverse momentum of the reconstructed candidate;
\item $\mid \eta^\mathrm{LAB}(\Lambda,\,\bar{\Lambda})\mid < 1.2$, where  $\eta^\mathrm{LAB}(\Lambda,\,\bar{\Lambda})$ 
is the pseudorapidity of the reconstructed candidate in the laboratory;
\item $M(e^+e^-) >  0.05 \gev$, 
to eliminate electron pairs from photon conversions
\footnote{$M(ab)$ is defined as the invariant mass for two vertex tracks 
with the assignment of masses of particles $a$ and $b$.}; 
\item $M(\pi^+\pi^-) < 0.475\gev$, to remove $K^{0}_{S}$ contamination;
\item $1.11 < M(p \pi) < 1.122 \gev$.
\end{itemize} 

The $K^{0}_{S}$ meson candidates were reconstructed from the decays to $\pi^+\pi^-$ (branching ratio 68.95 $\pm$ 0.14\% \cite{pl:b592:1}). Both tracks were assigned the mass of the charged pion. Additional requirements to select $K^{0}_{S}$ are given in the following:
\begin{itemize}
\item $0.6 < P^\mathrm{LAB}_T(K^{0}_{S}) < 2.5\gev$, where $P^\mathrm{LAB}_T(K^{0}_{S})$ is the transverse momentum of the reconstructed candidate;
\item $\mid \eta^\mathrm{LAB}(K^{0}_{S})\mid < 1.2$, where  $\eta^\mathrm{LAB}(K^{0}_{S})$ is the pseudorapidity of the reconstructed candidate in the laboratory;
\item $M(e^+e^-) >  0.05 \gev$; 
\item $M(p\pi) > 1.125 \gev$, to remove $\Lambda$ and $\bar{\Lambda}$ contamination. Here the mass of the proton was assigned to the track with larger momentum and the mass of the pion to the other track;
\item $0.48 < M(\pi^{+} \pi^{-}) < 0.52 \gev$.
\end{itemize} 

\begin{figure}[p]
\vfill
\begin{center}
\epsfig{file=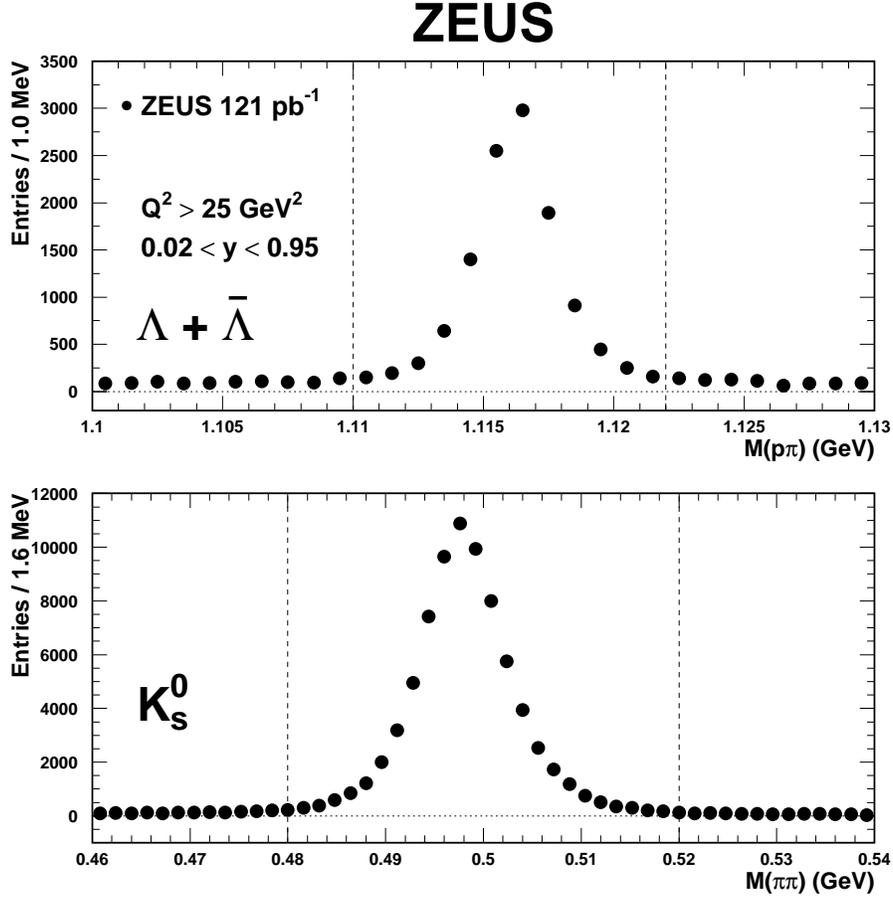,width=13cm}
\end{center}
\caption{
Mass distributions of the secondary vertex candidates in the $\Lambda + \bar{\Lambda}$ and $K^{0}_{S}$ samples
assuming $p \pi$ and $\pi \pi$ decays respectively.  
Only candidates in the range $0.6 < P_{T}^\mathrm{LAB} < 2.5 \gev$ and $|\eta^\mathrm{LAB}| < 1.2$ for events with $Q^{2} > 25 \gev^2$ 
and $0.02 < y < 0.95$ are displayed.  Statistical errors are smaller than the point size. The number of total $\Lambda + \bar{\Lambda}$ 
and $K^{0}_{S}$ candidates located within the vertical lines are estimated to be 10731 and 73140, respectively, after the background subtraction. 
}
\label{lamk0s_peaks_hiq2}
\vfill
\end{figure}

\begin{figure}[p]
\vfill
\begin{center}
\epsfig{file=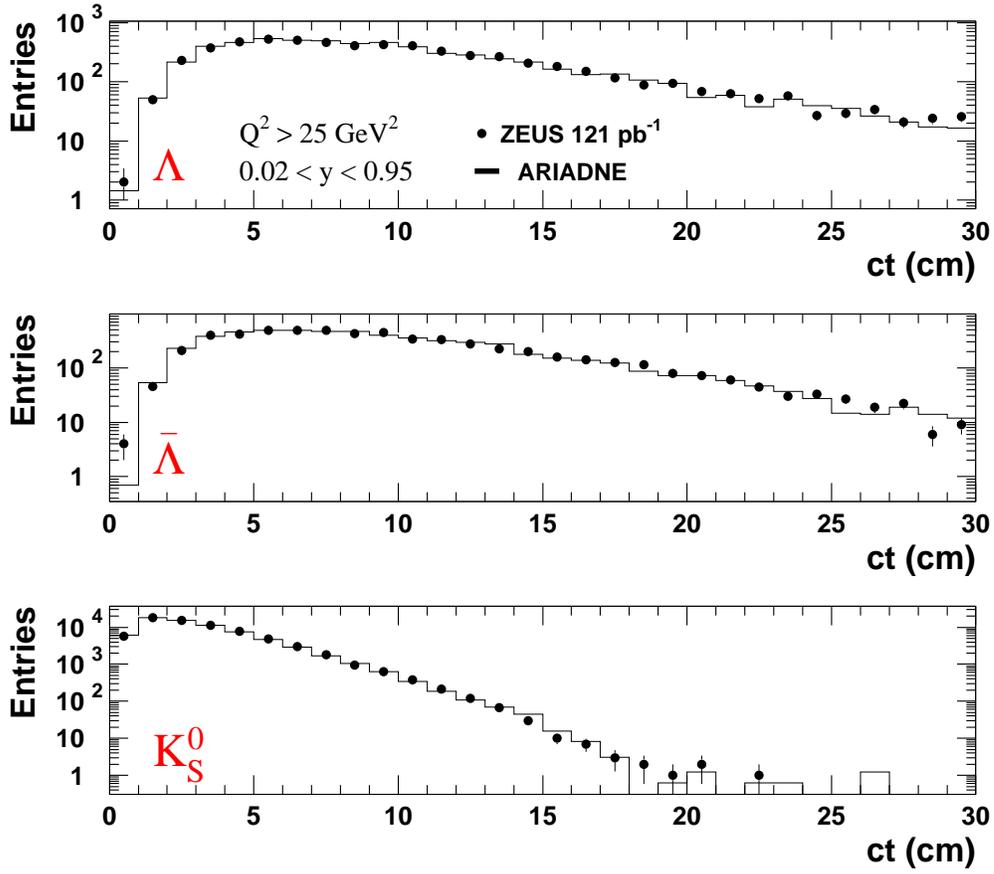,width=13cm}
\end{center}
\caption{
Detector-level distributions of $ct$, where $t$ is the reconstructed proper lifetime, 
for $\Lambda$, $\bar{\Lambda}$ and $K^{0}_{S}$ samples for data and {\sc Ariadne}.
The {\sc Ariadne} histogram is normalised to the same number of events as the data.  
Statistical errors are shown unless smaller than the point size.  
}
\label{lifetime_recon_hiq2}
\vfill
\end{figure}

The mass peaks for $K^{0}_{S}$ and $\Lambda + \bar{\Lambda}$ in the high-$Q^{2}$ sample are shown in Fig. \ref{lamk0s_peaks_hiq2}. The decay of $K^{0}_{S}$, $\Lambda$, and $\bar{\Lambda}$ was well understood as can be demonstrated by Fig. \ref{lifetime_recon_hiq2}, which shows the proper decay times, reconstructed from the three dimensional decay length, compared to the expectations from MC
simulation (see below).

\section{Event simulation}\label{s:mc}
Production of $K^{0}_{S}$, $\Lambda$ and $\bar{\Lambda}$ hadrons was modelled using the MC programs described below. 
In these models, strange quarks can be produced perturbatively by the boson-gluon fusion process 
($\gamma g \rightarrow s \bar{s}$) or by gluon splitting in so-called parton showers. They may also originate from 
the proton parton densities or can be generated in non-perturbative string fragmentation. Strange hadrons 
are produced during hadronization, when quarks recombine into hadrons, and through the decays of other hadrons. 
Samples of events were generated to determine the response of the detector and obtain the correction factors 
required to convert the detector-level distributions to the hadron level. The generated events were passed 
through a full simulation of the detector, using {\sc Geant} 3.13 \cite{tech:cern-dd-ee-84-1}, and processed 
with the same reconstruction program as used for the data.  

The high-$Q^{2}$ and the low-$Q^{2}$ DIS data were corrected to the hadron level using 
the {\sc Ariadne} 4 \cite{cpc:71:15} MC program interfaced to 
{\sc Heracles} 4.6.1 \cite{cpc:69:155,cpc:81:381,spi:www:heracles,spi:www:djangoh11} 
via {\sc Djangoh} 1.1\cite{spi:www:djangoh11}, to include QED corrections. 
The CTEQ proton parton density functions were used \cite{pr:d55:1280,epj:c12:375}. 
{\sc Ariadne} is based on the Colour Dipole Model in which most QCD coherence effects are 
modelled as gluon emission from colour dipoles between partons. The program uses the 
Lund string model\cite{prep:97:31} to simulate the fragmentation of the partons. 
A significant parameter governing the production of strange hadrons  is the strangeness-suppression factor, 
$\lambda_s$, that is probability to produce $s$-quark pairs relative 
to $u$- and $d$-quark pairs in the string fragmentation. This was set to 0.3, the default value,  
as found in $e^+ e^-$ annihilation \cite{zfp:c56:521,*zfp:c68:1,*zfp:c69:379,*zfp:c73:61}. 
Other parameters that control baryon production were set to their 
default values\footnote{The key parameters for baryon production in JETSET are the diquark-antidiquark 
pair production suppression PARJ(1)=0.10, the suppression of s quark pair production compared to u or d 
pair production $\lambda_s \equiv$ PARJ(2)=0.30, the extra suppression of strange diquark production compared with the normal 
suppression of strange quarks PARJ(3)=0.4 and the suppression of spin 1 diquarks compared with spin 0 ones PARJ(4)=0.05.}
\cite{hep-ph-0108264}. 
This {\sc Ariadne} sample was also used to compare to the final cross sections and ratios.  
 
A value of $\lambda_s$ smaller than 0.3 is often preferred \cite{zfp:c68:29,*epj:c2:77,*np:b480:3,*zfp:c76:213,zfp:c65:587,zfp:c61:539,*pr:d59:052001} for $K^{0}_{S}$ production. Therefore, a further sample with 
$\lambda_s = 0.22$ \cite{pl:b553:141} was also generated and used for comparison. The DIS data were additionally 
simulated using the {\sc Lepto} 6.5 MC program \cite{cpc:101:108}, which is based on first-order matrix elements 
plus parton showers (MEPS). The same Lund string model was used for the hadronisation, with $\lambda_s = 0.3$, 
and the same proton parton density functions
as in the {\sc Ariadne} 
sample. This was used for further comparison to the data.  

The photoproduction data were corrected to the hadron level using the {\sc Pythia} 6 event generator 
\cite{cpc:135:238}, which consists of leading-order matrix-element calculations with initial-  and 
final-state parton showering to simulate higher-order processes. The proton and photon PDFs were taken 
from GRV \cite{zfp:c67:433} and SaS2D \cite{zfp:c68:607} respectively. Multiple interactions
\cite{pr:d36:2019,*pl:b300:169,*np:b407:539,*jp:g19:1657,zfp:c70:17,epj:c1:109}, where more than one pair 
of partons (one parton from the photon and one parton from the proton) interact independently, were included. 
The default implementation was used, with the $p^\mathrm{min}_{T}$ \cite{hep-ph-0108264} value of $2.7 \gev$. 
The hadronisation is performed by the Lund string model, as in {\sc Ariadne}, with the same parameters 
controlling the production of strange hadrons. Direct and resolved events were generated separately. 
For correction of the data, the direct and resolved subsamples were combined such that they gave a best fit 
to the data $x_{\gamma}^\mathrm{OBS}$ distribution. The {\sc Pythia} sample was also used to compare to the 
final cross sections and ratios, in which case the direct and resolved events were combined according to the 
predicted cross sections.  

\section{Cross-section determination}

The cross sections in the high-$Q^{2}$ DIS sample were measured in the kinematic region $Q^{2} >25 \gev^2$ and $0.02 < y < 0.95$. The cross sections in the low-$Q^{2}$ DIS sample were measured in the kinematic region $5 < Q^{2} < 25 \gev^2$ and $0.02 < y < 0.95$. The cross sections in the photoproduction sample were measured in the kinematic region $Q^{2} < 1 \gev^2$ and $0.2 < y < 0.85$, with the additional requirement of 2 jets, both satisfying $E_{T}^{\mathrm{jet}} > 5 \gev$ and $|\eta^{\mathrm{jet}}|<2.4$. In all three samples there was a further kinematic requirement that $0.6 < P_T^{\mathrm{LAB}}(K^{0}_{S},\Lambda,\bar{\Lambda}) < 2.5 \gev$ and $|\eta^\mathrm{LAB}(K^{0}_{S},\Lambda,\bar{\Lambda})| < 1.2$. In all samples the measured cross sections were the luminosity-weighted average of the cross sections at the centre-of-mass energies $\sqrt{s} = 318 \gev$ and $\sqrt{s} = 300 \gev$. In the DIS samples differential cross sections were measured as functions of $P_T^\mathrm{LAB}(K^{0}_{S},\Lambda,\bar{\Lambda})$, $\eta^\mathrm{LAB}(K^{0}_{S},\Lambda,\bar{\Lambda})$, $x_\mathrm{Bj}$, and $Q^{2}$. In the photoproduction sample differential cross sections were measured as functions of $P_T^\mathrm{LAB}(K^{0}_{S},\Lambda,\bar{\Lambda})$, $\eta^\mathrm{LAB}(K^{0}_{S},\Lambda,\bar{\Lambda})$, and $x_{\gamma}^\mathrm{OBS}$.

The $K^{0}_{S}$, $\Lambda$ and $\bar{\Lambda}$ differential cross sections in any variable $Y$ were calculated
using a standard bin-by-bin correction 
as follows:
\begin{equation*} \label{cross_section_eqn}
\frac{d \sigma}{d Y} = \frac{N}{A \cdot \mathcal{L} \cdot B \cdot \Delta Y},
\end{equation*}
where $N$ is the number of $K^{0}_{S}$, $\Lambda$ or $\bar{\Lambda}$ in a bin of width $\Delta Y$, 
$\mathcal{L}$ is the luminosity, $A$ is the acceptance and $B$ is the branching ratio. 
The acceptance 
was calculated from the MC samples described in Section \ref{s:mc}, and took into account migration 
effects and efficiencies for each bin. The acceptance for each particle was calculated for each bin 
of a $6 \times 6$ grid in $p_{T}$ and $\eta$.
An additional acceptance correction was made to obtain the differential cross sections as functions 
of variables other than $p_{T}$ or $\eta$. In each bin a mass-sideband subtraction method was used to 
subtract the remaining combinatorial background in the $K^{0}_{S}$, $\Lambda$ and $\bar{\Lambda}$ samples, 
which was at the level of $\sim 3$\% in the $K^{0}_{S}$ sample and $\sim 6$\% in the $\Lambda$ and $\bar{\Lambda}$ samples.

The differential cross sections, $\frac{d \sigma}{d Y}$, for the considered particle
as a function of $\eta$ and $P_T$ can be converted into the multiplicity distribution 
$\frac{1}{d \sigma_{\mathrm{inc}}} \frac{d \sigma}{d Y}$, where $\sigma_{\mathrm{inc}}$ is the inclusive 
event cross section using the following factors: $\sigma_{\mathrm{inc}}^{-1}$ =  
$2.6\times10^{-5}\, \mathrm{pb}^{-1}$ for the high-$Q^2$ sample,
$8.9\times10^{-6}\, \mathrm{pb}^{-1}$ for the low-$Q^2$ sample
and $1.6\times10^{-5}\, \mathrm{pb}^{-1}$ for the photoproduction sample. 
On average, in the measured $P_T$ and $\eta$ region, for the high-$Q^2$ sample, there were about 0.017 
$\Lambda$ (or $\bar{\Lambda}$) and 0.09 $K^{0}_{S}$ per event. The corresponding multiplicities 
for the photoproduction sample were 0.077 for $\Lambda$ (or $\bar{\Lambda}$) and 0.27 for $K^{0}_{S}$.
\subsection{Systematic uncertainties}
The main sources which contributed to the systematic uncertainty were investigated by changing the analysis 
procedure, as outlined below, and observing the difference from the primary result. The total systematic 
uncertainty in the DIS samples for each bin was calculated by adding the individual contributions from the 
different variations in quadrature. The following sources of systematic uncertainties were considered: 
the energy measurement of the scattered electron;
the measurement of $Q^{2}$, $\delta$, $Z_{\mathrm{vtx}}$
and $y$;
measurements of secondary-track $p_{T}$ and $\eta$;
the removal of $\Lambda$ and $K^{0}_{S}$ due to the collinearity cut;
the impact position of the electron on the CAL;
the estimation of the background.
The systematic uncertainty on the cross sections was generally $\lesssim 5$\%.
Additionally, a 2\% uncertainty due to the luminosity measurement was included for the cross sections. 
The branching-ratio uncertainties were deemed negligible and not taken into account.

The most significant systematic error in the photoproduction sample was due to the uncertainty in 
the calorimeter energy scale ($\pm 3$\%), shown as a separate band in the figures. Typical uncertainties 
in the cross sections were $\lesssim 10$\%, except in the highest bin of $x_{\gamma}^\mathrm{OBS}$ where 
the uncertainty was up to $80$\%. A systematic uncertainty of 7\% on the photoproduction cross sections 
due to trigger efficiencies was included. The limitations of the {\sc Pythia} Monte Carlo simulation, 
particularly in describing the $x_{\gamma}^\mathrm{OBS}$ distribution in the data, introduced a possible additional 
systematic uncertainty in correcting the data from detector level to hadron level. The effect of this was 
investigated by reweighting the {\sc Pythia} $x_{\gamma}^\mathrm{OBS}$ distribution to the data and using 
this reweighted sample to correct the data. The difference between the results obtained with the reweighted 
sample compared to the primary results was treated as a further systematic uncertainty; uncertainties in 
the cross sections were $\lesssim 5$\%.

The uncertainty of the tracking simulation \cite{epj:c44:351} was negligible
compared to all other sources.
Most of the uncertainties discussed above cancel in the ratios and asymmetries
presented in this paper.
\section{Results and discussion}
\subsection{Cross sections}

\begin{figure}[p]
\vfill
\begin{center}
\epsfig{file=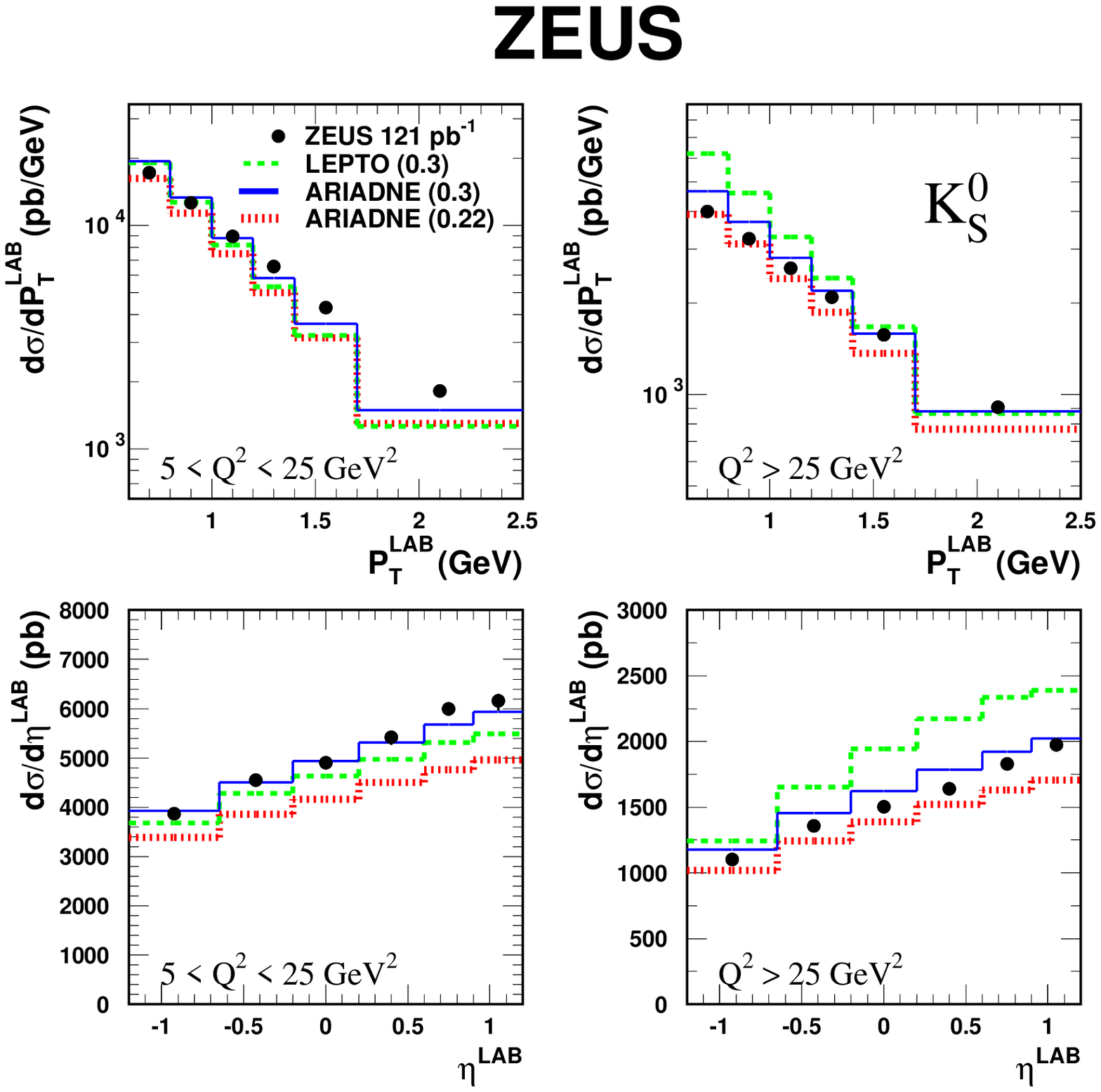,width=13cm}
\end{center}
\caption{
Differential $K^{0}_{S}$ cross-sections as a function of $P_{T}^\mathrm{LAB}$ and $\eta^\mathrm{LAB}$
in the range $0.6 < P_{T}^\mathrm{LAB} < 2.5 \gev$ and $|\eta^\mathrm{LAB}| < 1.2$ for events with $5 < Q^{2} < 25 \gev^2$, 
$0.02 < y < 0.95$ and $Q^{2} > 25 \gev^2$, 
$0.02 < y < 0.95$.  Statistical errors (inner error bars) and the systematic uncertainties added in 
quadrature are shown, unless smaller than the point size.  
The histograms show predictions from {\sc Ariadne} and {\sc Lepto} using the stated strangeness suppression.
}
\label{k0s_xsec_dis_pteta}
\vfill
\end{figure}

\begin{figure}[p]
\vfill
\begin{center}
\epsfig{file=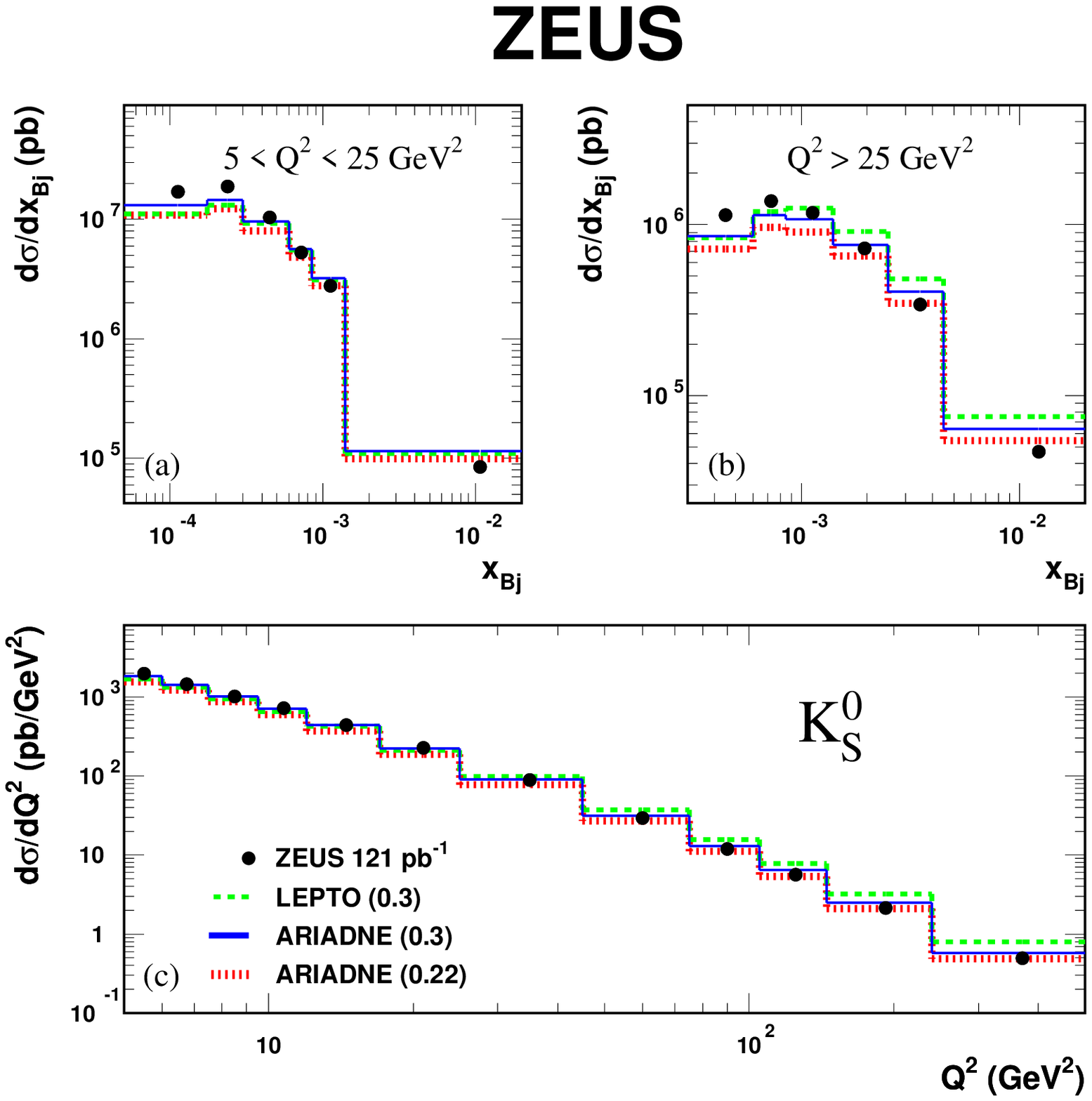,width=13cm}
\end{center}
\caption{
Differential $K^{0}_{S}$ cross sections as a function of $x_\mathrm{Bj}$ and $Q^{2}$
in the range $0.6 < P_{T}^\mathrm{LAB} < 2.5 \gev$ and $|\eta^\mathrm{LAB}| < 1.2$ for events 
a) with $5 < Q^{2} < 25 \gev^2$, 
$0.02 < y < 0.95$ b) $Q^{2} > 25 \gev^2$, 
$0.02 < y < 0.95$ and c) $Q^{2} > 5 \gev^2$, $0.02 < y < 0.95$.  
Statistical errors and the systematic uncertainties added in 
quadrature are smaller than the point size.  
The histograms show predictions from {\sc Ariadne} and {\sc Lepto} using the stated strangeness suppression.
}
\label{k0s_xsec_dis_xq2}
\vfill
\end{figure}

\begin{figure}[p]
\vfill
\begin{center}
\epsfig{file=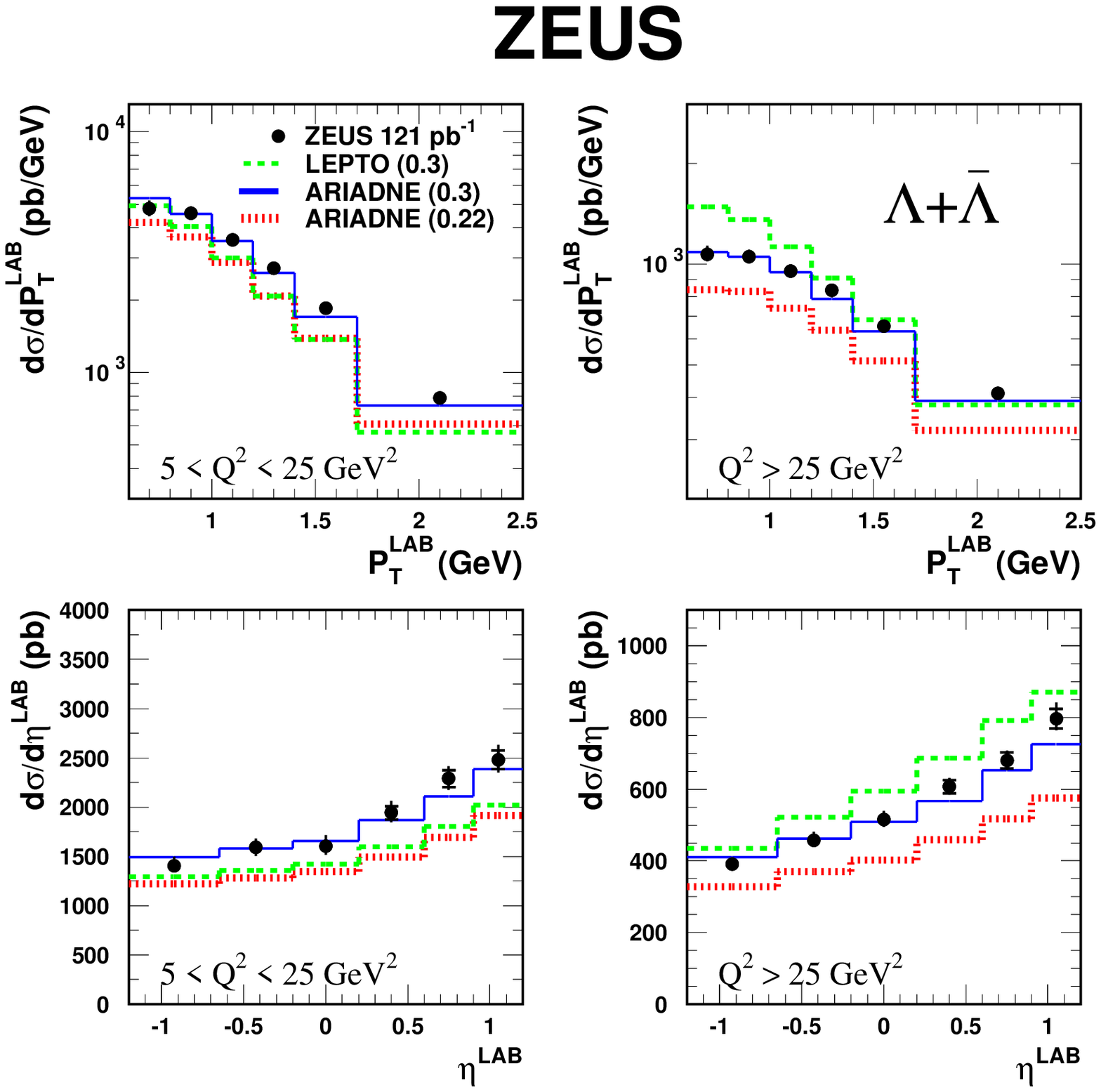,width=13cm}
\end{center}
\caption{
Differential $\Lambda + \bar{\Lambda}$ cross sections as a function of $P_{T}^\mathrm{LAB}$ and $\eta^\mathrm{LAB}$
in the range $0.6 < P_{T}^\mathrm{LAB} < 2.5 \gev$ and $|\eta^\mathrm{LAB}| < 1.2$ for events with $5 < Q^{2} < 25 \gev^2$, 
$0.02 < y < 0.95$ and $Q^{2} > 25 \gev^2$, 
$0.02 < y < 0.95$.  Statistical errors (inner error bars) and the systematic uncertainties added in 
quadrature are shown, unless smaller than the point size.  
The histograms show predictions from {\sc Ariadne} and {\sc Lepto} using the stated strangeness suppression.
}
\label{lamalm_xsec_dis_pteta}
\vfill
\end{figure}

\begin{figure}[p]
\vfill
\begin{center}
\epsfig{file=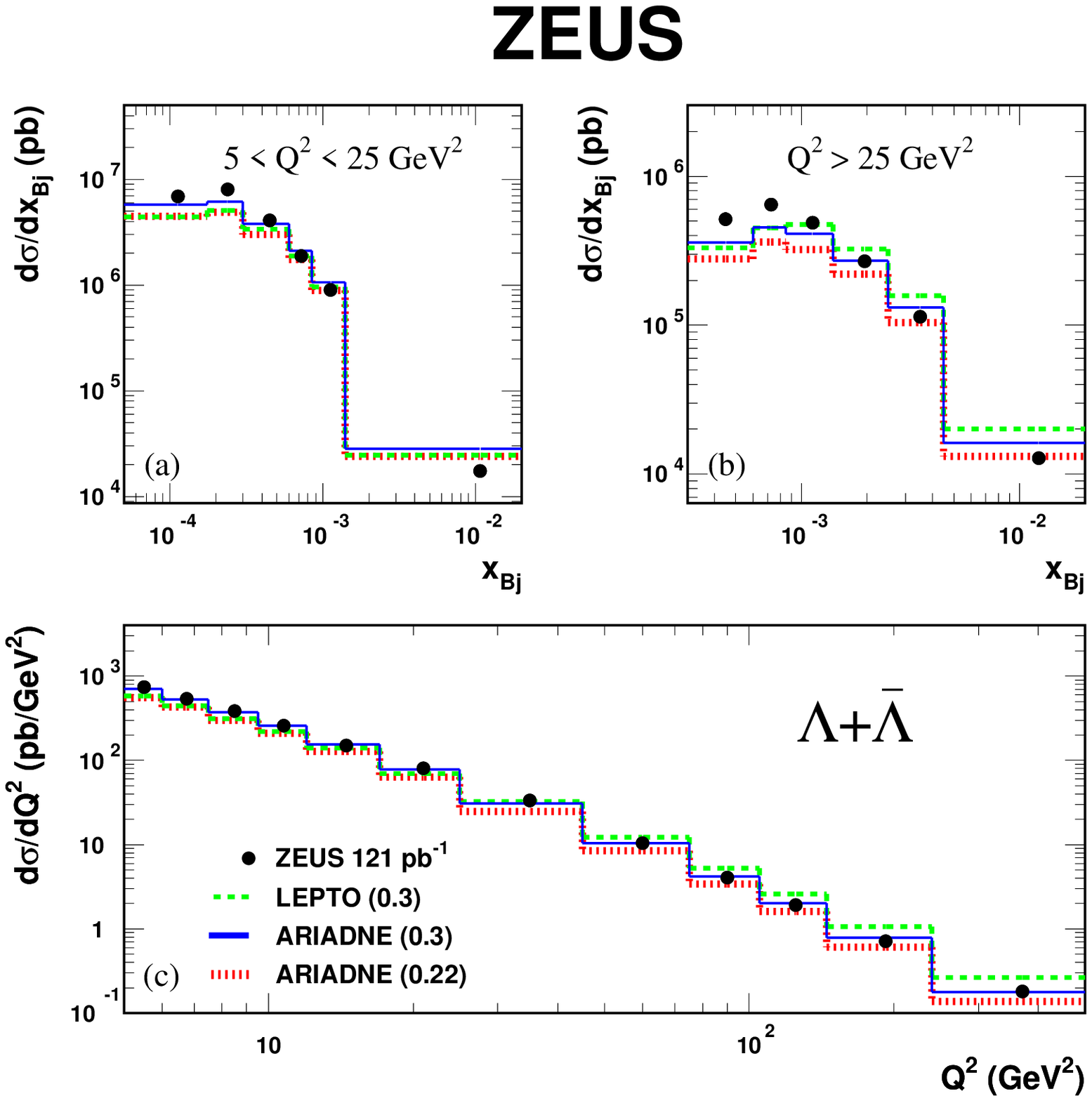,width=13cm}
\end{center}
\caption{
Differential $\Lambda + \bar{\Lambda}$ cross sections as a function of $x_\mathrm{Bj}$ and $Q^{2}$
in the range $0.6 < P_{T}^\mathrm{LAB} < 2.5 \gev$ and $|\eta^\mathrm{LAB}| < 1.2$ for events 
a) with $5 < Q^{2} < 25 \gev^2$, 
$0.02 < y < 0.95$ b) $Q^{2} > 25 \gev^2$, 
$0.02 < y < 0.95$ and c) $Q^{2} > 5 \gev^2$, $0.02 < y < 0.95$. 
Statistical errors and the systematic uncertainties added in 
quadrature are smaller than the point size. 
The histograms show predictions from {\sc Ariadne} and {\sc Lepto} using the stated strangeness suppression. 
}
\label{lamalm_xsec_dis_xq2}
\vfill
\end{figure}

\begin{figure}[p]
\vfill
\begin{center}
\epsfig{file=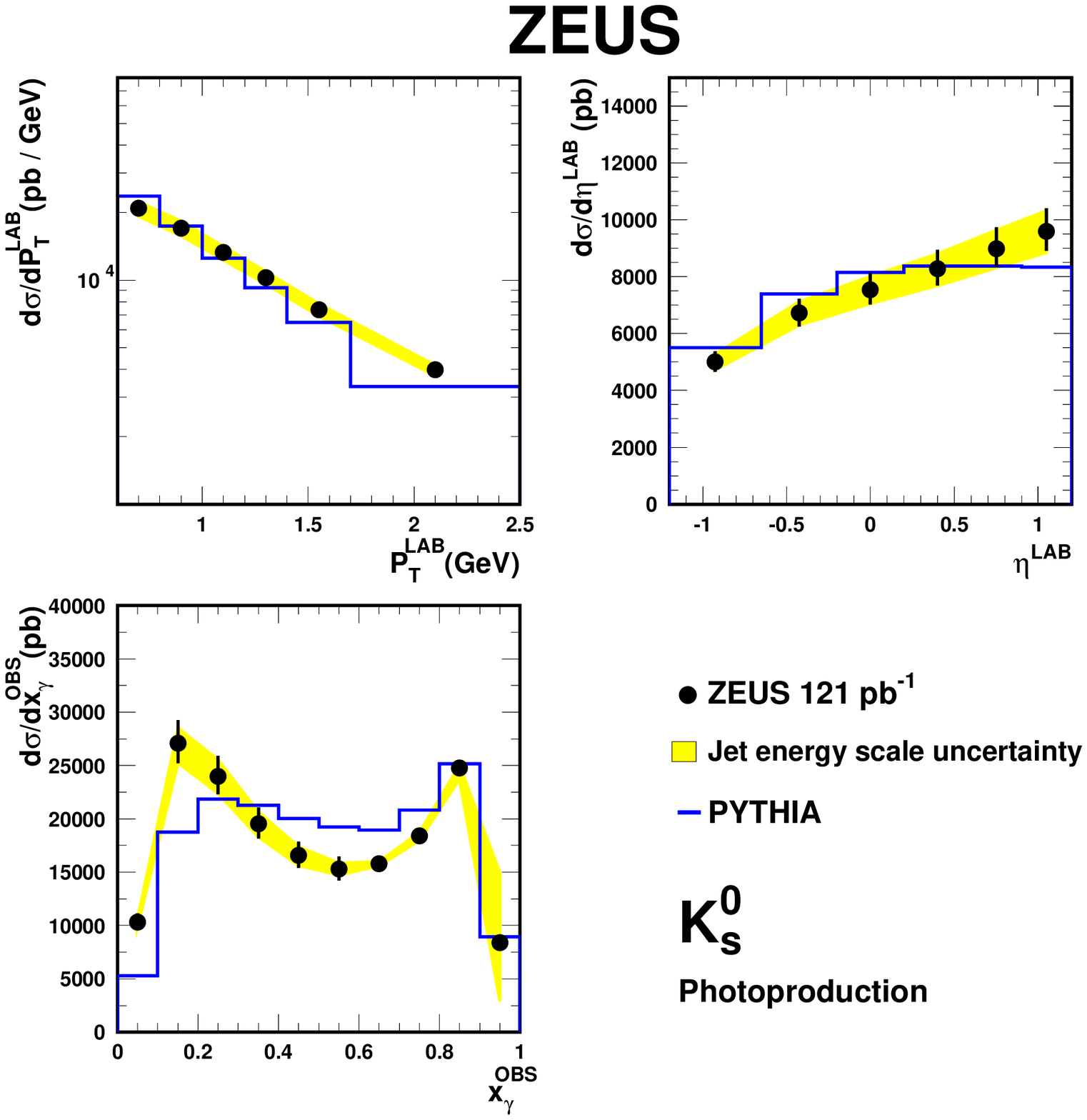,width=13cm}
\end{center}
\caption{
Differential $K^{0}_{S}$ cross sections as a function of $P_{T}^\mathrm{LAB}$, $\eta^\mathrm{LAB}$ 
and $x_{\gamma}^\mathrm{OBS}$, in the range $0.6 < P_{T}^\mathrm{LAB} < 2.5 \gev$ and $|\eta^\mathrm{LAB}| < 1.2$ for 
events with $Q^{2} < 1 \gev^2$, $0.2 < y < 0.85$ and at least two jets both satisfying $E_{T}^{\mathrm{jet}} > 5 \gev$
and $|\eta^{\mathrm{jet}}|<2.4$.  
Statistical errors are shown, unless smaller than the point size, together with the 
systematic uncertainty arising from the trigger efficiency added in quadrature.  
The uncertainty arising from the jet energy scale is also shown (shaded band). 
The solid histogram shows the prediction from {\sc Pythia} (with multiple interactions), normalised to the data. 
}
\label{k0s_xsec_php}
\vfill
\end{figure}

\begin{figure}[p]
\vfill
\begin{center}
\epsfig{file=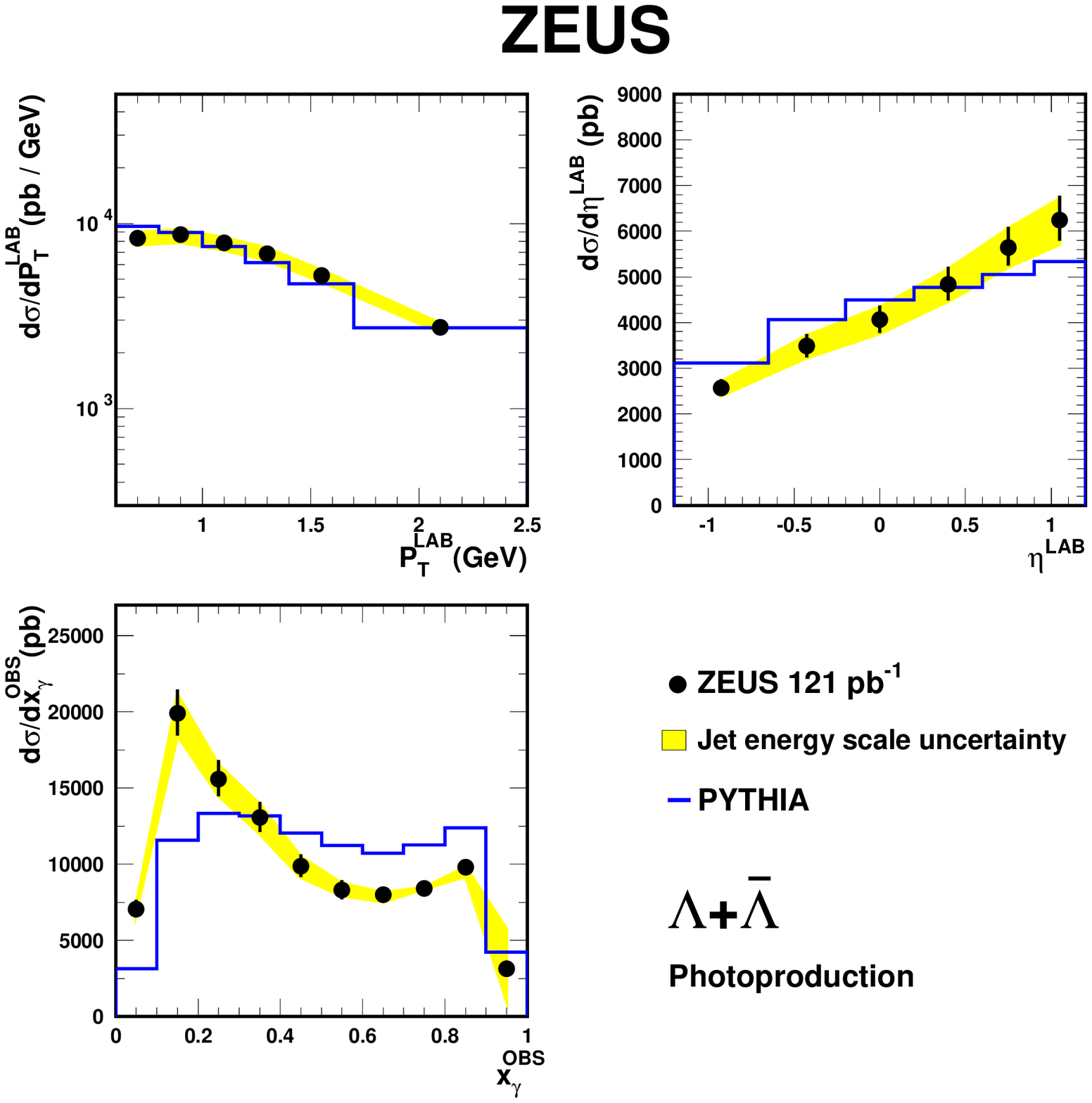,width=13cm}
\end{center}
\caption{
Differential $\Lambda + \bar{\Lambda}$ cross sections as a function of $P_{T}^\mathrm{LAB}$, $\eta^\mathrm{LAB}$ 
and $x_{\gamma}^\mathrm{OBS}$, in the range $0.6 < P_{T}^\mathrm{LAB} < 2.5 \gev$ and $|\eta^\mathrm{LAB}| < 1.2$ for 
events with $Q^{2} < 1 \gev^2$, $0.2 < y < 0.85$ and at least two jets both satisfying $E_{T}^{\mathrm{jet}} > 5 \gev$
and $|\eta^{\mathrm{jet}}|<2.4$.  
Statistical errors are shown, unless smaller than the point size, together with the 
systematic uncertainty arising from the trigger efficiency added in quadrature.  
The uncertainty arising from the jet energy scale is also shown (shaded band). 
The solid histogram shows the prediction from {\sc Pythia} (with multiple interactions), normalised to the data. 
}
\label{lamalm_xsec_php}
\vfill
\end{figure}

Measured differential cross sections for the production of $K^{0}_{S}$ and $\Lambda$ + $\bar{\Lambda}$ are shown in 
Figs. \ref{k0s_xsec_dis_pteta} - \ref{lamalm_xsec_dis_xq2} for the DIS data and in  
Figs. \ref{k0s_xsec_php} and \ref{lamalm_xsec_php} for the photoproduction sample.  
The DIS cross sections are compared to the absolute predictions of {\sc Ariadne} and {\sc Lepto} MC calculations.  
The photoproduction cross sections are compared to the prediction of the {\sc Pythia} MC with multiple 
interactions normalised to the data cross section.  

The {\sc Ariadne} program with strangeness-suppression factor of 0.3 describes the data reasonably well, although the 
$K^{0}_{S}$ cross section for the high-$Q^2$ sample is overestimated. The slope of the $P_{T}^\mathrm{LAB}$ 
dependence is incorrect and the cross section at low $x_\mathrm{Bj}$ is underestimated for both low- and 
high-$Q^2$ samples. A similar comment can be made for the $\Lambda$ + $\bar{\Lambda}$ cross sections.

The description of the data by {\sc Ariadne} with the strangeness-suppression factor of 0.22 is less satisfactory 
but, as $\lambda_{s}$ is not the only parameter influencing the cross section, a conclusion
on what value of the $\lambda_s$ can describe the data best can not be drawn. 
The {\sc Lepto} MC does not describe 
the data well.

In photoproduction, {\sc Pythia} with multiple interactions describes the shape of the data dependence on 
$P_{T}^\mathrm{LAB}$ and $\eta^\mathrm{LAB}$ adequately, but the $x_{\gamma}^\mathrm{OBS}$ dependence is 
predicted to be too flat and at the smallest $x_{\gamma}^\mathrm{OBS}$, in the resolved photon region, 
the description is poor.

\subsection{Baryon-antibaryon asymmetry}

The baryon-antibaryon asymmetry $\mathcal{A}$ is defined as:
\begin{equation*} \label{baryon_asymmetry_eqn}
\mathcal{A} = \frac{N(\Lambda) - N(\bar{\Lambda})}{N(\Lambda) + N(\bar{\Lambda})},
\end{equation*}
where $N(\Lambda)$, $N(\bar{\Lambda})$ is the number of $\Lambda$ and $\bar{\Lambda}$
baryons, respectively.  

The baryon-antibaryon asymmetry $\mathcal{A}$ has been measured and compared to MC
predictions from {\sc Ariadne} and {\sc Pythia} 
(with $\lambda_{s} = 0.3$ for both, and also with $\lambda_{s} = 0.22$ for {\sc Ariadne}). 
The following values were obtained:
\begin{itemize}
\item at high $Q^2$: $\mathcal{A}$ = $0.3 \pm 1.3^{+0.5}_{-0.8} $\%, compared to 
the {\sc Ariadne} ($\lambda_{s} = 0.3$) prediction of $0.4 \pm 0.2 $\%;
\item at low $Q^2$: $\mathcal{A}$ = $1.2 \pm 1.6^{+0.7}_{-2.1}$\%, compared to 
the {\sc Ariadne} ($\lambda_{s} = 0.3$) prediction of $1.0 \pm 0.2 $\%;
\item in photoproduction: $\mathcal{A}$ = $-0.07 \pm 0.6^{+1.0}_{-1.0} $\%, compared to
the {\sc Pythia} prediction of $0.6 \pm 0.1 $\%.
\end{itemize}

\begin{figure}[p]
\vfill
\begin{center}
\epsfig{file=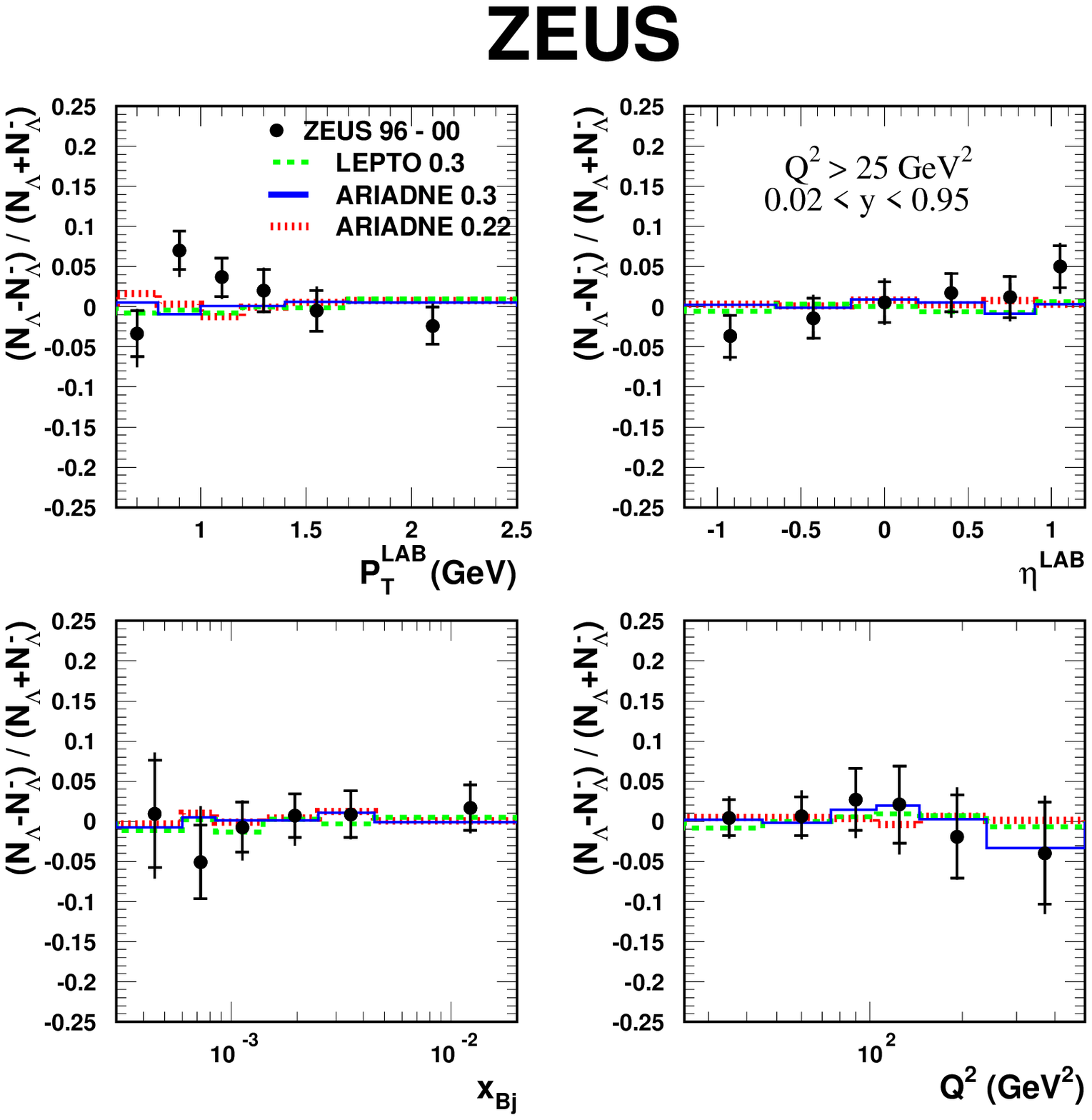,width=13cm}
\end{center}
\caption{
The ratio $\frac{N(\Lambda) - N(\bar{\Lambda})}{N(\Lambda) + N(\bar{\Lambda})}$ as a function of 
$P_{T}^\mathrm{LAB}$, $\eta^\mathrm{LAB}$, $x_\mathrm{Bj}$ and $Q^{2}$, 
in the range $0.6 < P_{T}^\mathrm{LAB} < 2.5 \gev$ and $|\eta^\mathrm{LAB}| < 1.2$ for events with $Q^{2} > 25 \gev^2$ 
and $0.02 < y < 0.95$.  
Statistical errors (inner error bars) and the systematic uncertainties added in quadrature are shown.  
The histograms show predictions from {\sc Ariadne} and {\sc Lepto} using the stated strangeness suppression.
}
\label{lam_asym_hiq2}
\vfill
\end{figure}

\begin{figure}[p]
\vfill
\begin{center}
\epsfig{file=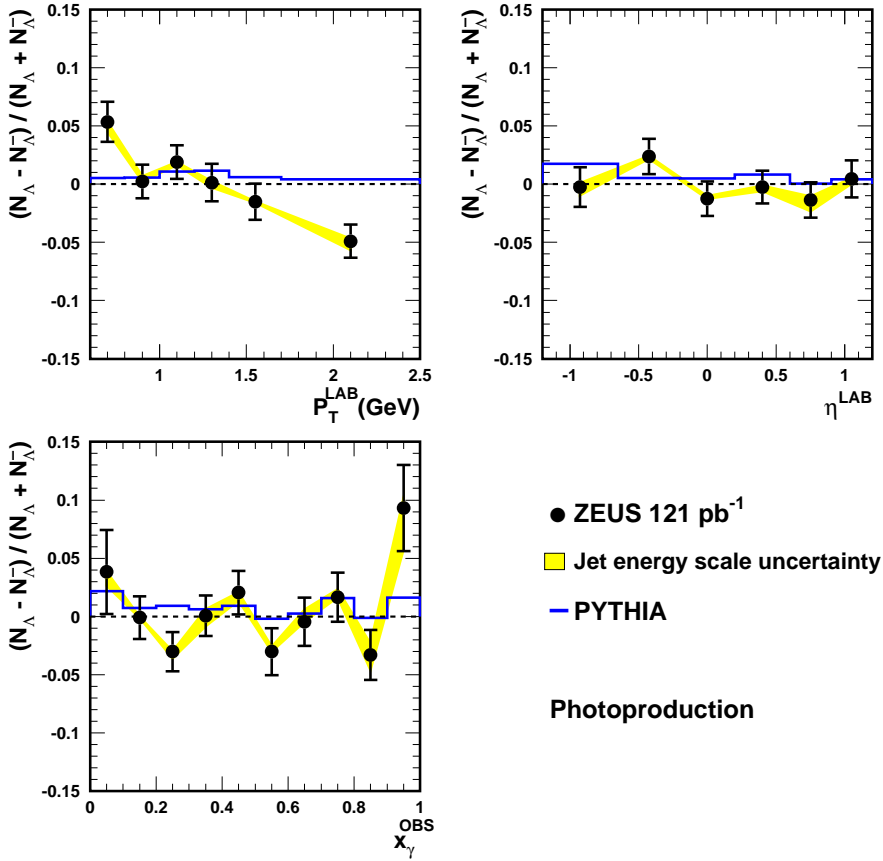,width=13cm}
\end{center}
\caption{
The ratio $\frac{N(\Lambda) - N(\bar{\Lambda})}{N(\Lambda) + N(\bar{\Lambda})}$ as a function of 
$P_{T}^\mathrm{LAB}$, $\eta^\mathrm{LAB}$ 
and $x_{\gamma}^\mathrm{OBS}$, in the range $0.6 < P_{T}^\mathrm{LAB} < 2.5 \gev$ and $|\eta^\mathrm{LAB}| < 1.2$ for 
events with $Q^{2} < 1 \gev^2$, $0.2 < y < 0.85$ and at least two jets both satisfying $E_{T}^{\mathrm{jet}} > 5 \gev$
and $|\eta^{\mathrm{jet}}|<2.4$.  
Statistical errors are shown, together with the uncertainty arising 
from the jet energy scale (shaded band). 
The solid histogram shows the prediction from {\sc Pythia} (with multiple interactions). 
}
\label{lam_asym_php}
\vfill
\end{figure}

Figures \ref{lam_asym_hiq2} and \ref{lam_asym_php} show the baryon-antibaryon asymmetry at high-$Q^2$
as a function of $P_{T}^\mathrm{LAB}$, $\eta^\mathrm{LAB}$, $x_\mathrm{Bj}$, $Q^{2}$
and in photoproduction as a function of $P_{T}^\mathrm{LAB}$, $\eta^\mathrm{LAB}$ and $x_{\gamma}^\mathrm{OBS}$.
In all cases, the average baryon-antibaryon asymmetry is consistent both with no asymmetry and consistent with 
the very small asymmetry predicted by Monte Carlo. This suggests that in the considered parts of the $ep$ 
phase-space, to a good approximation, baryons and antibaryons are produced according to the same mechanism. 

A positive asymmetry of 3.5\% is predicted in DIS \cite{zfp:c75:693}, due to the so called gluon-junction mechanism 
that makes it possible for the ``baryon number to travel" several units of rapidity, in this case from the proton 
beam direction to the rapidity around 0 in the laboratory frame in which the measurements were made. Such a class 
of models can describe the significant positive baryon-antibaryon asymmetry which has been measured at 
the heavy-ion collider RHIC \cite{prl:89:092302}. Although this prediction is not $Q^2$ dependent, it is 
not clear whether it could be extended down to $Q^2 = 0$ and applied to the selected photoproduction sample. 
Combining statistical and systematic errors from the three samples, the average asymmetry is in disagreement 
with the 3.5\% value.

Although {\sc Ariadne} predicts that on average about 15\% (40\% at the highest $x_\mathrm{Bj}$ for 
the low-$Q^2$ sample) of events
with a reconstructed $\Lambda$ or $\bar{\Lambda}$ originate from the exchanged photon coupling to 
an $s$ or $\bar{s}$ quark from the proton, this measurement, being at low $x_\mathrm{Bj}$, is not 
sensitive to the strange-quark asymmetry in the proton structure function as studied by NuTeV \cite{pr:d:2006}.

As our baryon-antibaryon asymmetry is consistent with no asymmetry, the $\Lambda$ and $\bar{\Lambda}$
samples were combined together (except for the transverse polarization measurement) 
and results presented for the combined sample. 

\subsection{Baryon-to-meson ratio}

The baryon-to-meson ratio $\mathcal{R}$ is defined as:
\begin{equation*} \label{baryon_ratio_eqn}
\mathcal{R} = \frac{N(\Lambda) + N(\bar{\Lambda})}{N(K^{0}_{S})},
\end{equation*}
where $N(\Lambda), N(\bar{\Lambda}), N(K^{0}_{S})$ refer to the number of indicated
hadrons.

\begin{figure}[p]
\vfill
\begin{center}
\epsfig{file=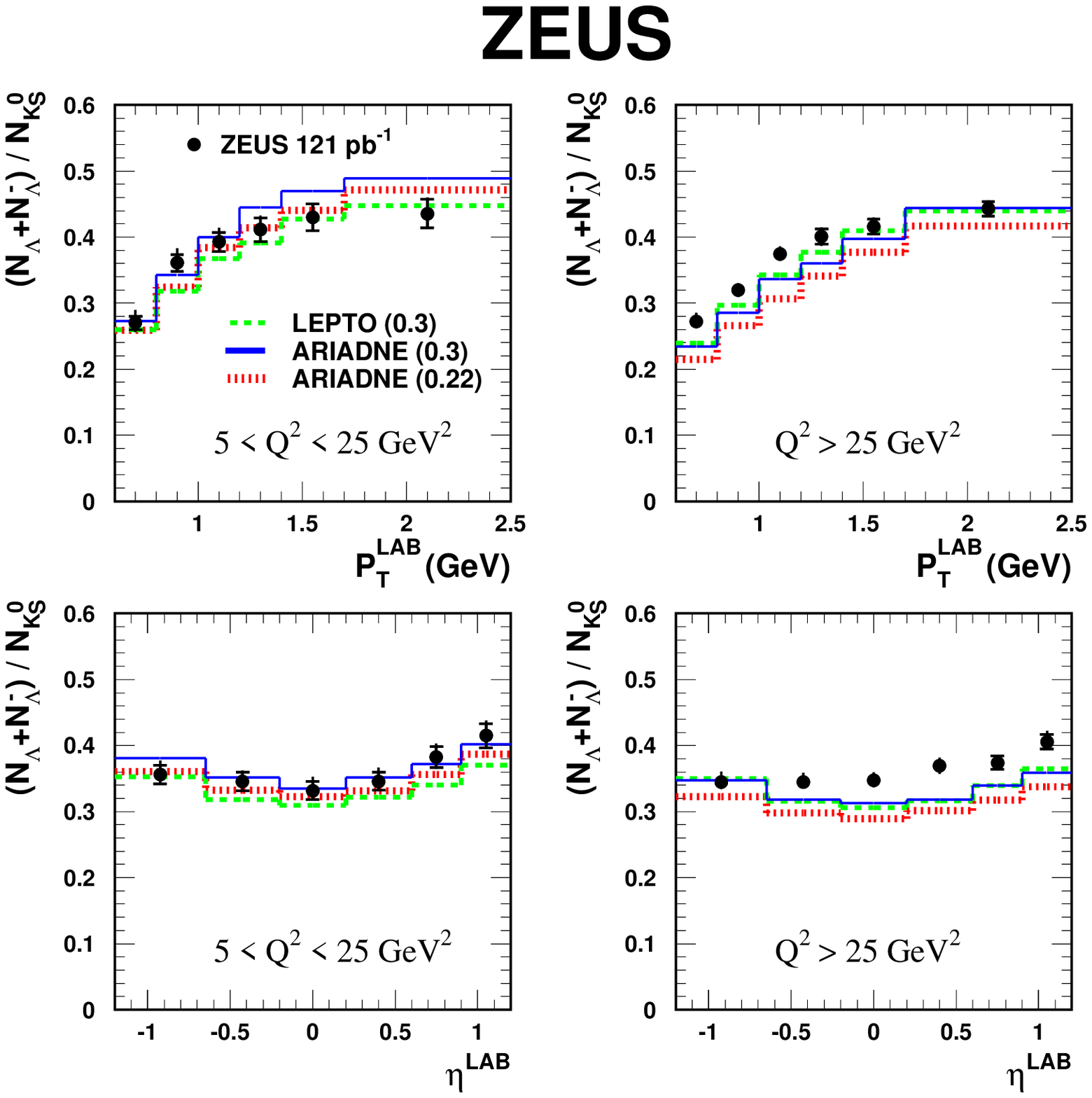,width=13cm}
\end{center}
\caption{
The ratio $\frac{N(\Lambda) + N(\bar{\Lambda})}{N(K^{0}_{S})}$ as a function of 
$P_{T}^\mathrm{LAB}$ and $\eta^\mathrm{LAB}$
in the range $0.6 < P_{T}^\mathrm{LAB} < 2.5 \gev$ and $|\eta^\mathrm{LAB}| < 1.2$ for events with $5 < Q^{2} < 25 \gev^2$, 
$0.02 < y < 0.95$ and $Q^{2} > 25 \gev^2$, 
$0.02 < y < 0.95$.  
Statistical errors (inner error bars) and the systematic uncertainties added in quadrature are shown.  
The histograms show predictions from {\sc Ariadne} and {\sc Lepto} using the stated strangeness suppression.
}
\label{lamk0s_ratio_hiq2}
\vfill
\end{figure}

\begin{figure}[p]
\vfill
\begin{center}
\epsfig{file=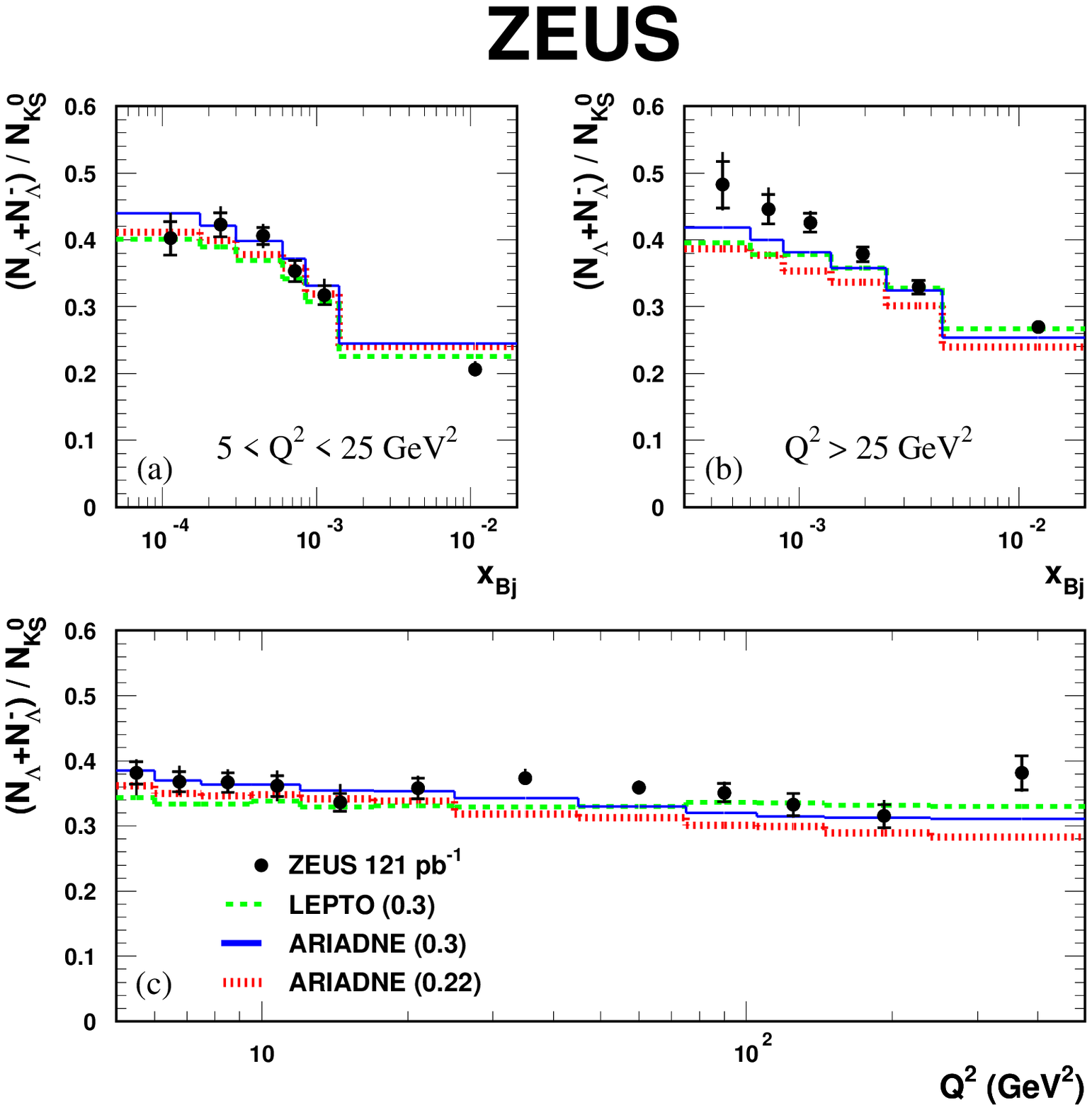,width=13cm}
\end{center}
\caption{
The ratio $\frac{N(\Lambda) + N(\bar{\Lambda})}{N(K^{0}_{S})}$ as a function of 
$x_\mathrm{Bj}$ and $Q^{2}$
in the range $0.6 < P_{T}^\mathrm{LAB} < 2.5 \gev$ and $|\eta^\mathrm{LAB}| < 1.2$ for events with 
a) $5 < Q^{2} < 25 \gev^2$, 
$0.02 < y < 0.95$ b) $Q^{2} > 25 \gev^2$, 
$0.02 < y < 0.95$ and c) $Q^{2} > 5 \gev^2$, $0.02 < y < 0.95$.  
Statistical errors (inner error bars) and the systematic uncertainties added in quadrature are shown. 
The histograms show predictions from {\sc Ariadne} and {\sc Lepto} using the stated strangeness suppression. 
}
\label{lamk0s_ratio_loq2}
\vfill
\end{figure}

\begin{figure}[p]
\vfill
\begin{center}
\epsfig{file=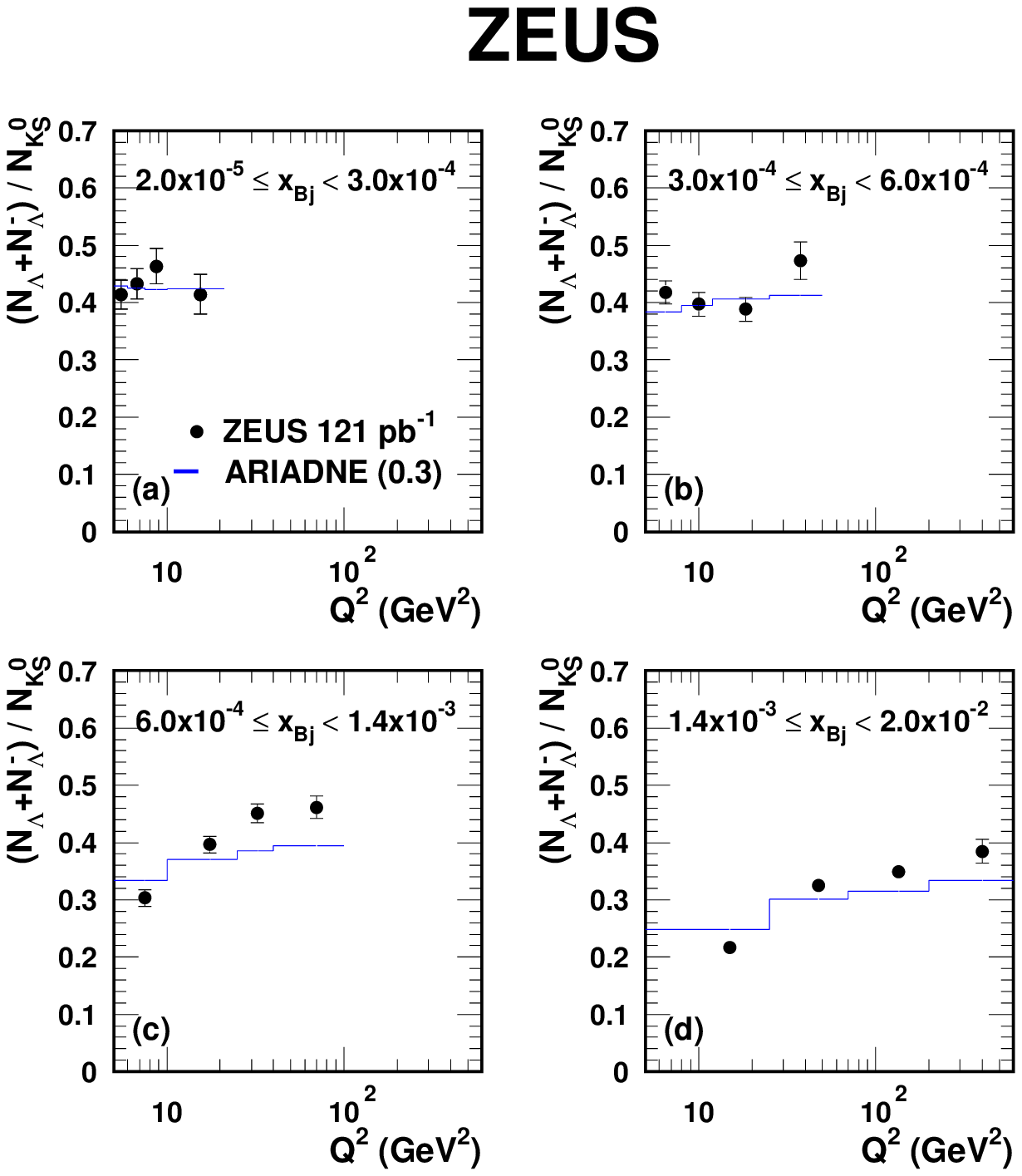,width=12cm}
\end{center}
\caption{
The ratio $\frac{N(\Lambda) + N(\bar{\Lambda})}{N(K^{0}_{S})}$ as a function of 
$Q^{2}$ for four bins of $x_\mathrm{Bj}$, 
in the range $0.6 < P_{T}^\mathrm{LAB} < 2.5 \gev$ and $|\eta^\mathrm{LAB}| < 1.2$ for events with $Q^{2} > 5 \gev^2$ 
and $0.02 < y < 0.95$.  
Statistical errors only are shown.  
The histograms show predictions from {\sc Ariadne} with a strangeness-suppression factor of 0.3.
 }
\label{lamk0s_ratio_xvsq2}
\vfill
\end{figure}

\begin{figure}[p]
\vfill
\begin{center}
\epsfig{file=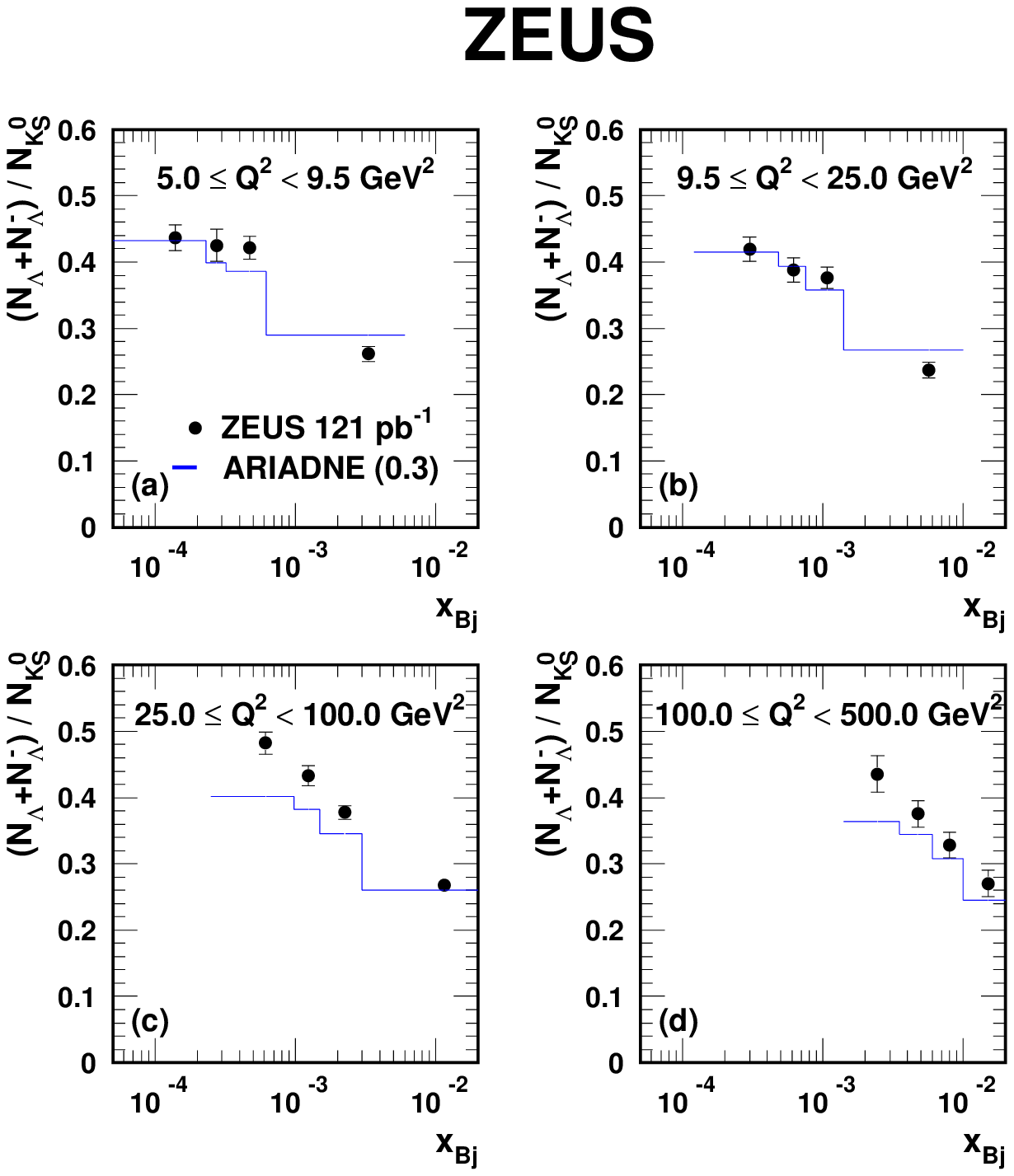,width=12cm}
\end{center}
\caption{
The ratio $\frac{N(\Lambda) + N(\bar{\Lambda})}{N(K^{0}_{S})}$ as a function of 
$x_\mathrm{Bj}$ for four bins of $Q^{2}$, 
in the range $0.6 < P_{T}^\mathrm{LAB} < 2.5 \gev$ and $|\eta^\mathrm{LAB}| < 1.2$ for events with $Q^{2} > 5 \gev^2$ 
and $0.02 < y < 0.95$.  
Statistical errors only are shown.  
The histograms show predictions from {\sc Ariadne} with a strangeness-suppression factor of 0.3.
}
\label{lamk0s_ratio_q2vsx}
\vfill
\end{figure}

Figures \ref{lamk0s_ratio_hiq2} and \ref{lamk0s_ratio_loq2} show the measured and predicted $\mathcal{R}$ 
for the DIS samples as a function of $P_{T}^\mathrm{LAB}$, $\eta^\mathrm{LAB}$, $x_\mathrm{Bj}$ 
and $Q^{2}$. The inaccuracies in describing the $K^{0}_{S}$ and $\Lambda$ + $\bar{\Lambda}$ cross sections 
for high $Q^{2}$, mentioned earlier, are clearly reflected in $\mathcal{R}$, but overall the {\sc Ariadne} 
MC with $\lambda_s = 0.3$ follows the shape of the data distributions and is usually in agreement 
to better than about 10\%. The low-$Q^2$ sample is described by the same Monte Carlo programs with even better accuracy. In order 
to have a better understanding of how $\mathcal{R}$ depends on $x_\mathrm{Bj}$ and $Q^{2}$, $\mathcal{R}$ 
is shown as a function of $Q^{2}$ for fixed bins in $x_\mathrm{Bj}$ and as a function of $x_\mathrm{Bj}$ 
for fixed bins in $Q^2$ in Figs. \ref{lamk0s_ratio_xvsq2} and \ref{lamk0s_ratio_q2vsx}. A dependence 
on $Q^2$ and a discrepancy between the data and MC can now be seen for the bins of higher $x_\mathrm{Bj}$. 
For the two bins of higher $Q^2$, the MC underestimates the data at low $x_\mathrm{Bj}$ by up to 20\%. 
The $\mathcal{R}$ value varies between about 0.2 and 0.5, and is about 0.4 to 0.5 at low $x_\mathrm{Bj}$ 
and low $Q^{2}$. These values can be compared to measurements at $e^+ e^-$  colliders, where
for centre-of-mass energies from 10 to 200 $\gev$, $\mathcal{R}$ varies between about 0.2 and 0.4 \cite{pl:b592:1}. 

\begin{figure}[p]
\vfill
\begin{center}
\epsfig{file=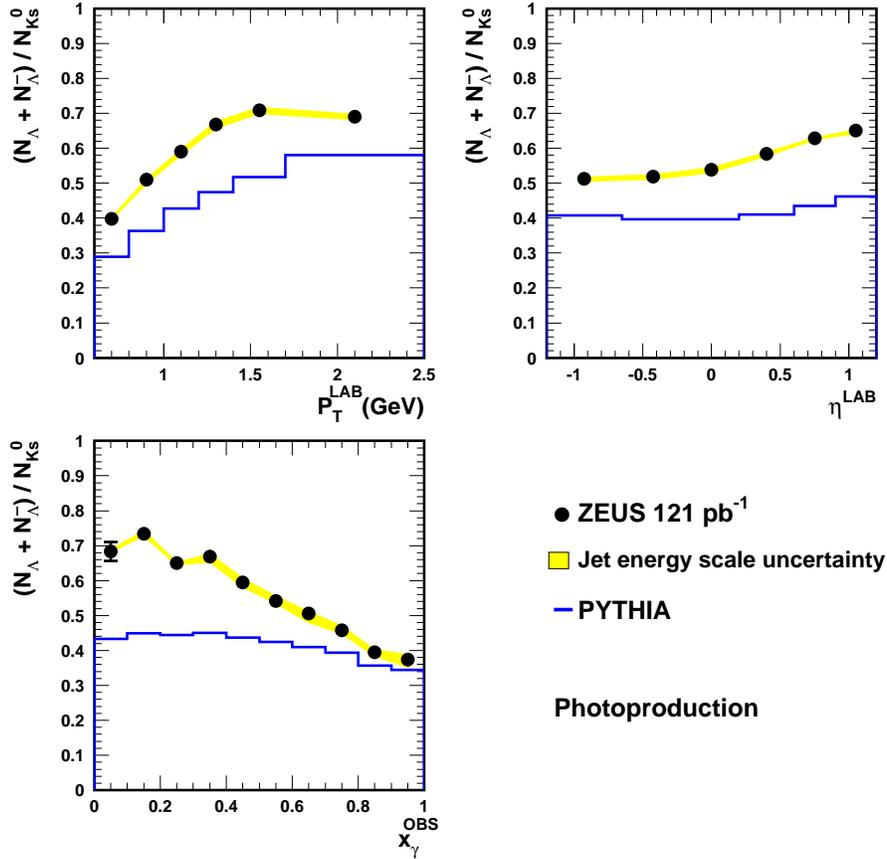,width=13cm}
\end{center}
\caption{
The ratio $\frac{N(\Lambda) + N(\bar{\Lambda})}{N(K^{0}_{S})}$ as a function of 
$P_{T}^\mathrm{LAB}$, $\eta^\mathrm{LAB}$ 
and $x_{\gamma}^\mathrm{OBS}$, in the range $0.6 < P_{T}^\mathrm{LAB} < 2.5 \gev$ and $|\eta^\mathrm{LAB}| < 1.2$ for 
events with $Q^{2} < 1 \gev^2$, $0.2 < y < 0.85$ and at least two jets both satisfying $E_{T}^{\mathrm{jet}} > 5 \gev$
and $|\eta^{\mathrm{jet}}|<2.4$.  
The statistical errors are shown, unless smaller than the point size.  
The shaded band shows the uncertainty arising from the jet energy scale. 
The solid histogram shows the prediction from {\sc Pythia} (with multiple interactions). 
}
\label{lamk0s_ratio_php}
\vfill
\end{figure}

Figure \ref{lamk0s_ratio_php} shows $\mathcal{R}$ for the photoproduction sample. For the direct-enriched sample, 
where  $x_{\gamma}^\mathrm{OBS} >$ 0.75, $\mathcal{R}$ is about 0.4, the same value as in DIS at low $x_\mathrm{Bj}$ 
and low $Q^{2}$.  
However, $\mathcal{R}$ rises to a value of about 0.7 towards low $x_{\gamma}^\mathrm{OBS}$ 
(resolved-enriched sample), while it stays flat in the {\sc Pythia} prediction.

\begin{figure}[p]
\vfill
\begin{center}
\epsfig{file=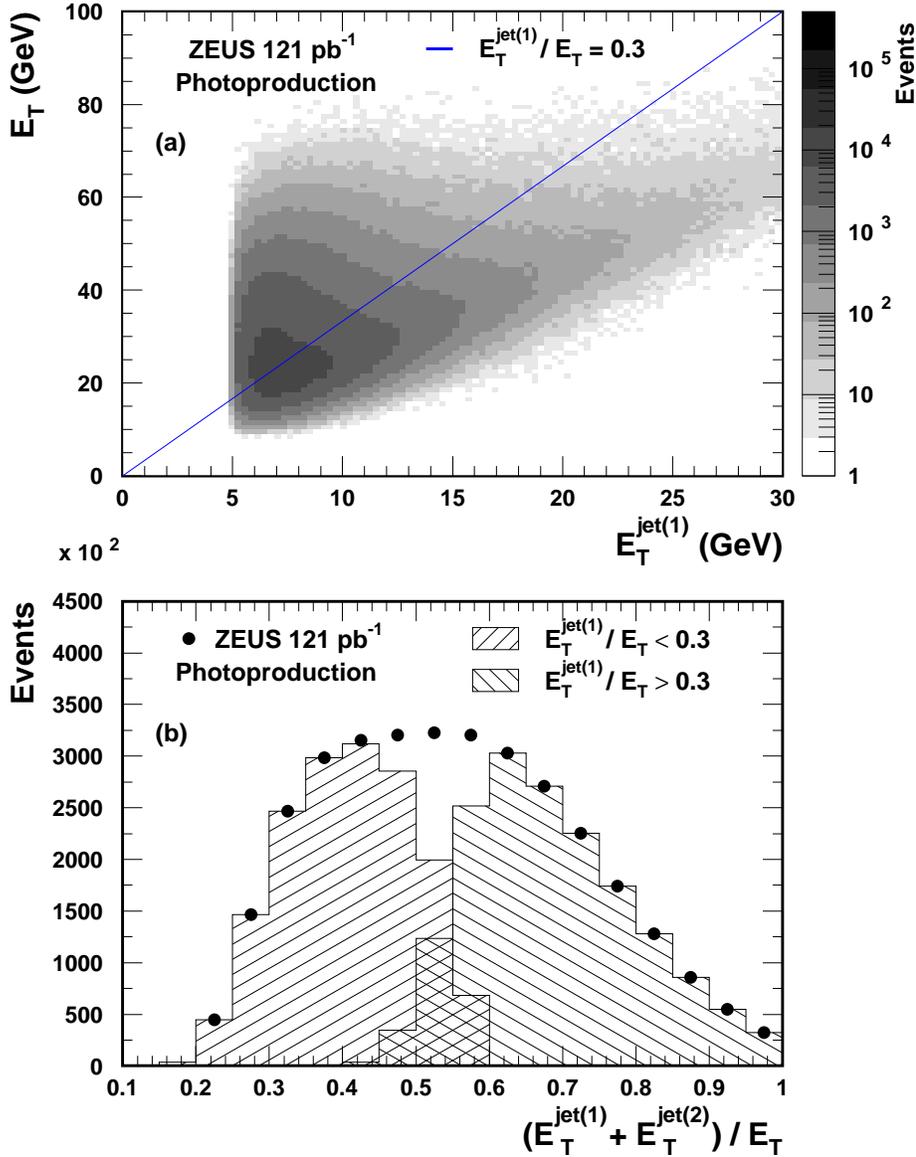,width=12cm}
\end{center}
\caption{
(a) Distribution of photoproduction events as a function of
the total transverse energy and the transverse energy of the jet with the highest transverse energy.
Here the event sample was selected without any strange hadron requirements.
The solid line represents the cut used to separate fireball-enriched and fireball-depleted subsamples.  
(b) The fraction of the total $E_{T}$ carried by the two jets of highest transverse energy for the same data sample as in (a).  
Fireball-enriched ($E_T^\mathrm{jet(1)}/E_T < 0.3$) and fireball-depleted ($E_T^\mathrm{jet(1)}/E_T > 0.3$) samples are shown.  Statistical errors are smaller than the point size.  
}
\label{jetet_vs_totalet_php}
\vfill
\end{figure}

In order to study this effect further, the photoproduction events were divided into two samples. In the first, 
the jet with the highest transverse energy was required to contribute at most 30\% to the total hadronic 
transverse energy. In this sample the events have largely isotropic transverse energy flow and therefore 
the sample is called ``fireball-enriched". The other sample, containing all the other events, was called ``fireball-depleted". Figure \ref{jetet_vs_totalet_php}(a) shows the distribution of events as a function of 
the total transverse energy, which is on average $30 \gev$, and of the transverse energy of the jet with 
the highest $E_{T}^{\mathrm{jet}}$. The line represents the cut used to separate fireball-enriched and 
fireball-depleted samples.
Figure \ref{jetet_vs_totalet_php}(b) illustrates the fireball selection in relation to the fraction of 
the total transverse energy carried by two jets of the highest transverse energy. It can be seen that 
fireball-depleted events are dominated by dijet events carrying most of the total transverse energy and 
that the fireball-enriched and fireball-depleted samples have about the same number of events.

\begin{figure}[p]
\vfill
\begin{center}
\epsfig{file=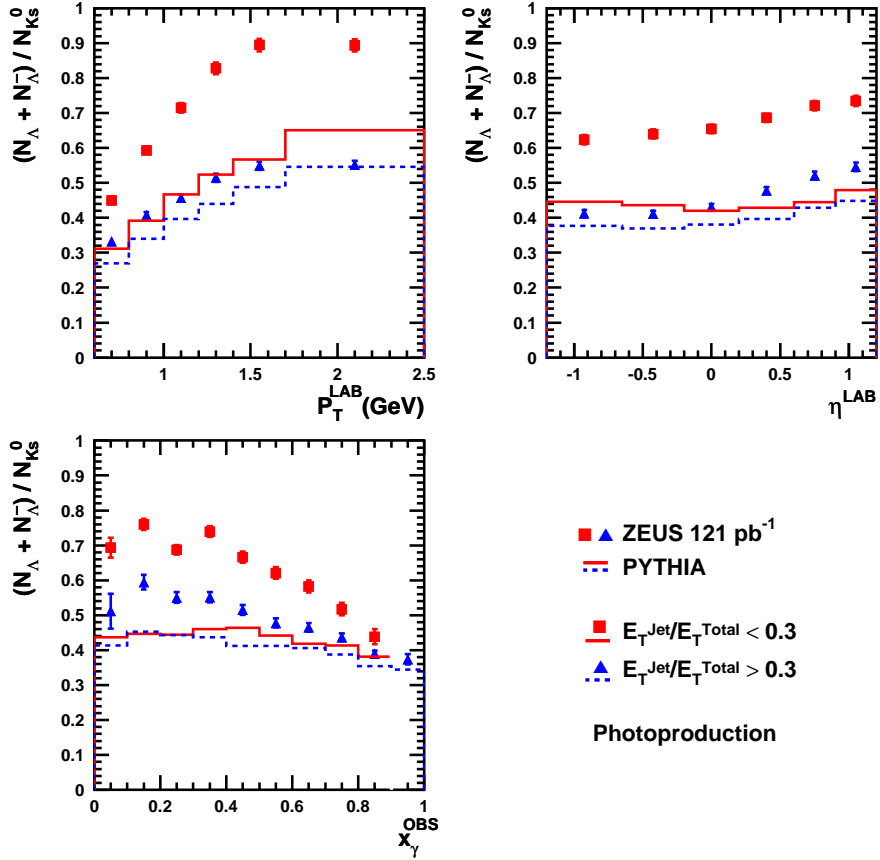,width=13cm}
\end{center}
\caption{
The ratio $\frac{N(\Lambda) + N(\bar{\Lambda})}{N(K^{0}_{S})}$ as a function of 
$P_{T}^\mathrm{LAB}$, $\eta^\mathrm{LAB}$ 
and $x_{\gamma}^\mathrm{OBS}$, in the range $0.6 < P_{T}^\mathrm{LAB} < 2.5 \gev$ and $|\eta^\mathrm{LAB}| < 1.2$ for 
events with $Q^{2} < 1 \gev^2$, $0.2 < y < 0.85$ and at least two jets both satisfying $E_{T}^{\mathrm{jet}} > 5 \gev$
and $|\eta^{\mathrm{jet}}|<2.4$.  
The ratios from the fireball-enriched sample (squares) and the fireball-depleted sample (triangles) 
are shown for the data.  The prediction from {\sc Pythia} for the fireball-enriched (solid line) and for the fireball-depleted (dashed line) samples
are shown.  Statistical errors are shown. 
The highest $x_{\gamma}^\mathrm{OBS}$ bin ($0.9 < x_{\gamma}^\mathrm{OBS} < 1.0$) of the fireball enriched sample is omitted due to 
insufficient statistics.  
}
\label{lamk0s_ratio_php_jetet}
\vfill
\end{figure}

The baryon-to-meson ratios for the fireball-enriched and fireball-depleted samples are presented 
in Fig. \ref{lamk0s_ratio_php_jetet} for the data and {\sc Pythia} MC. The measured $\mathcal{R}$ 
is larger for the fireball-enriched sample, most significantly at high $P_{T}^\mathrm{LAB}$, than it is 
for the fireball-depleted sample. This feature is not reproduced by {\sc Pythia}, which predicts almost 
the same $\mathcal{R}$ for both samples. The {\sc Pythia} prediction reasonably describes the measured 
values of $\mathcal{R}$ for the fireball-depleted sample. 
This is not surprising as {\sc Pythia} generates jets in events according to the multiple interaction mechanism \cite{pr:d36:2019,*pl:b300:169,*np:b407:539,*jp:g19:1657}, which makes several independent jets, like those in DIS or $e^+ e^-$ 
where baryons and mesons are created locally. Provided there is enough energy available, $\mathcal{R}$ will be the same, regardless of the number of jets (ignoring some differences in quark and gluon fragmentation).  


Large values of $\mathcal{R}$, larger than 1, have been measured
at hadron and heavy-ion colliders: $p \bar{p}$ \cite{pl:b366:441}, $pp$ \cite{nucl-ex-0607033,pl:b637:161} and
RHIC \cite{pr:c69:034909,prl:89:092301,prl:92:112301,pl:b595:143}. 
\subsection{Ratio of strange-to-light hadrons}

The ratio of strange-to-light hadrons was measured in terms of $\mathcal{T}$:
\begin{equation*} \label{to track_ratio_eqn}
\mathcal{T} = \frac{N(K^{0}_{S})}{N_\mathrm{ch}},
\end{equation*}
where $N(K^{0}_{S})$ is the number of $K^{0}_{S}$ and $N_\mathrm{ch}$ is the number 
of charged pions, charged kaons, protons and antiprotons,  
(excluding products of $K^{0}_{S}$, $\Lambda$, and $\bar{\Lambda}$ decays)
in the same region of $P_{T}^\mathrm{LAB}$ and $\eta^\mathrm{LAB}$ as the $K^{0}_{S}$.

\begin{figure}[p]
\vfill
\begin{center}
\epsfig{file=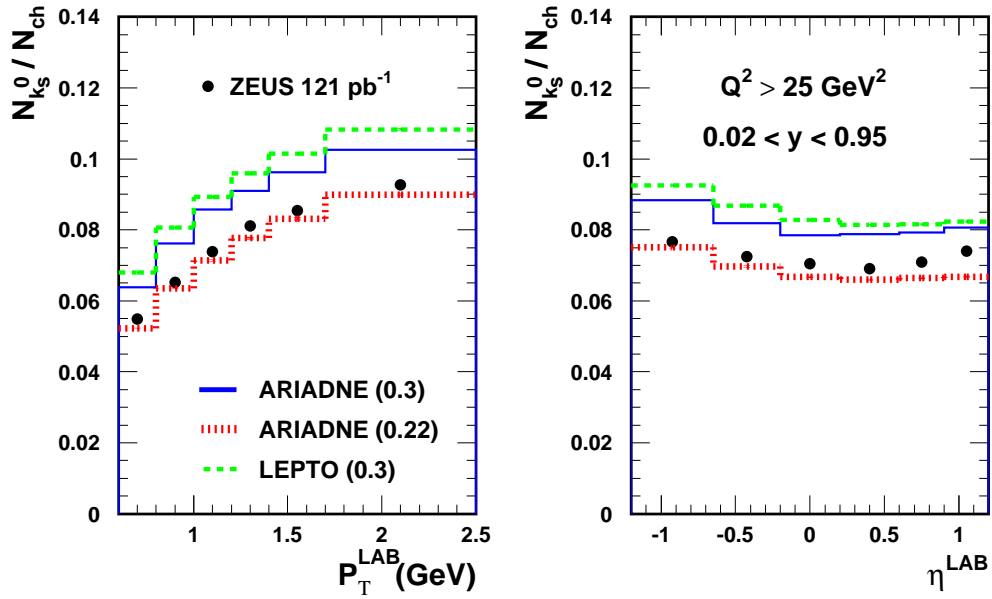,width=13cm}
\end{center}
\caption{
The ratio $\frac{N(K^{0}_{S})}{N_\mathrm{ch}}$ as a function of $P_{T}^\mathrm{LAB}$ and $\eta^\mathrm{LAB}$, 
in the range $0.6 < P_{T}^\mathrm{LAB} < 2.5 \gev$ and $|\eta^\mathrm{LAB}| < 1.2$ for events with $Q^{2} > 25 \gev^2$ 
and $0.02 < y < 0.95$.  
Statistical errors are smaller than the point size.  
The histograms show predictions from {\sc Ariadne} and {\sc Lepto} using the stated strangeness suppression.
}
\label{trk_ratio_hiq2}
\vfill
\end{figure}

\begin{figure}
\vfill
\begin{center}
\epsfig{file=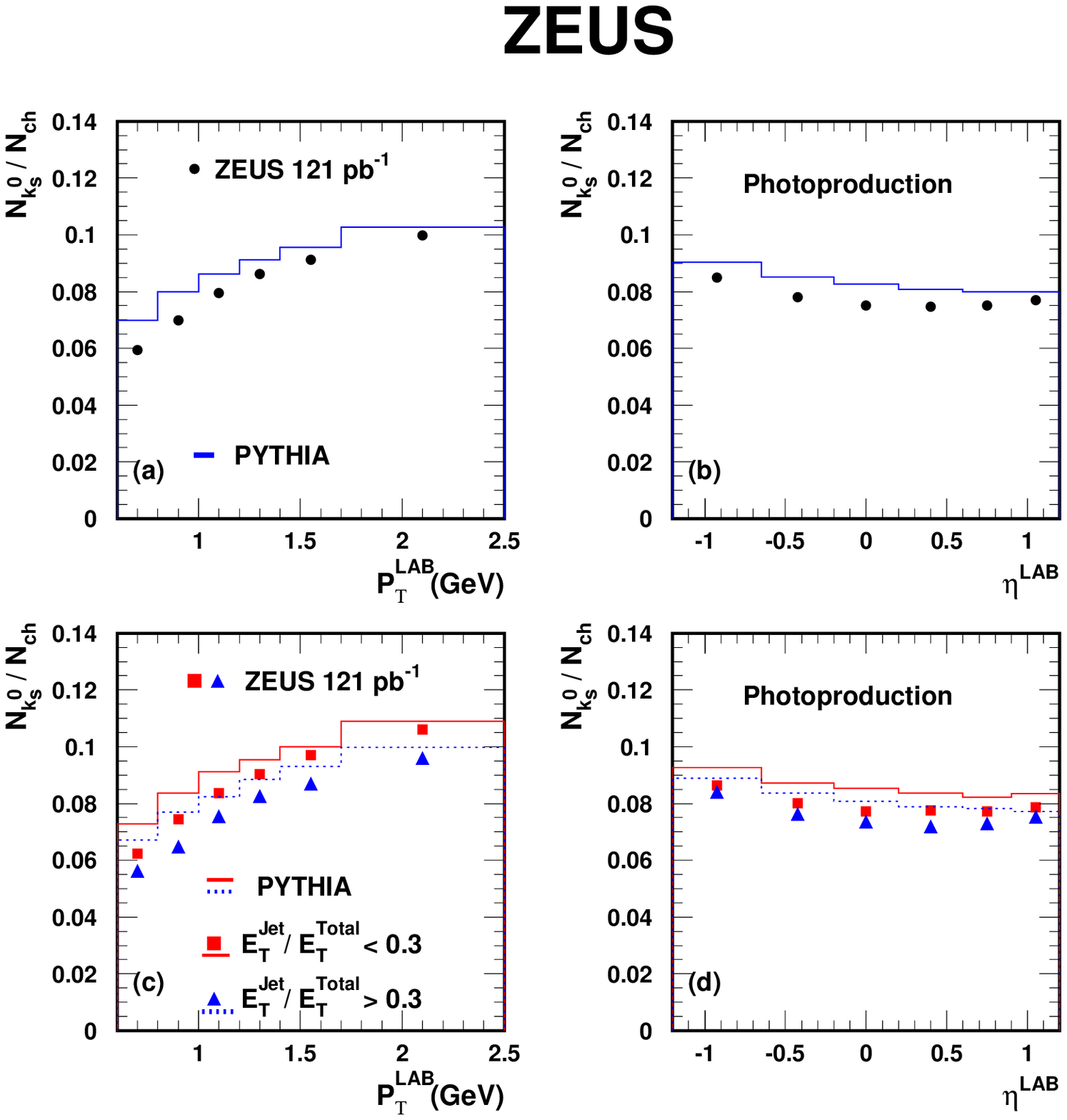,width=13cm}
\end{center}
\caption{
The ratio $\frac{N(K^{0}_{S})}{N_\mathrm{ch}}$ as a function of $P_{T}^\mathrm{LAB}$ and $\eta^\mathrm{LAB}$, 
in the range $0.6 < P_{T}^\mathrm{LAB} < 2.5 \gev$ and $|\eta^\mathrm{LAB}| < 1.2$ for events with $Q^{2} < 1 \gev^2$, 
$0.2 < y < 0.85$ and at least two jets both satisfying $E_{T}^{\mathrm{jet}} > 5 \gev$ and $|\eta^{\mathrm{jet}}|<2.4$. 
The ratio is shown for all events in (a) and (b) and for the fireball-enriched sample and the fireball-depleted sample 
in (c) and (d). Prediction from {\sc Pythia} (with multiple interactions) predictions with a strangeness suppression factor of 0.3 
are shown as solid and dashed histograms. Statistical errors are smaller than the symbols. 
}
\label{trk_ratio_php}
\vfill
\end{figure}

In Figs. \ref{trk_ratio_hiq2} and \ref{trk_ratio_php}, $\mathcal{T}$ is shown as a function of 
$P_{T}^\mathrm{LAB}$ and $\eta^\mathrm{LAB}$ for the high-$Q^2$ sample 
(for the low-$Q^2$ sample, not shown, the values are similar)
and for the photoproduction sample. 
The MC predictions from {\sc Ariadne} and {\sc Pythia} are also shown. They follow the data reasonably well, 
preferring the strangeness-suppression factor to be smaller than 0.3.  The measured $\mathcal{T}$ 
lies between 0.05 and 0.1, varying with $P_{T}^\mathrm{LAB}$ for both the DIS and photoproduction. Similar values have been measured 
at $e^+ e^-$ \cite{pl:b592:1} for the ratio of the number of $K^{0}_{S}$ to the number of charged 
pions and are on average about 0.07 at centre-of-mass energies from 10 to $35 \gev$, about 0.06 
at the $Z^{0}$ and about 0.05 at $200 \gev$. It can be concluded that $\mathcal{T}$ is about the same 
in $e^+ e^-$ and $ep$.

In order to see whether $\mathcal{T}$ depends on the transverse energy flow, the fireball selection, as discussed above, was applied to the photoproduction events. Figures \ref{trk_ratio_php}(c) and (d) show the measured and predicted $\mathcal{T}$
for the fireball selection. The quantity $\mathcal{T}$ hardly depends on the fireball selection, as predicted by {\sc Pythia}. 

Fireball events are candidates for events where QCD instantons 
\cite{pl:b59:85,prl:37:8,*pr:d14:3432,*pr:d18:2199,*prep:142:357,pl:b188:506,*np:b330:1,*np:b343:310,
*np:b365:3,*zfp:c66:285,*pl:b555:227} could 
play a role \cite{hep-ph-9411217,jp:g25:1440,hep-ph-9906441},  
since they are characterised by isotropic transverse energy flow.
Another expectation is a likely enhancement of heavier-quark production
relative to light quarks, due to the required flavour democracy. 
Searches for QCD instantons in DIS have been reported by H1 \cite{epj:c25:495}
and ZEUS \cite{epj:c34:255}. No effect was identified due to QCD instantons,
as the expected effects were small compared to the background at the relatively large $Q^2$ required. 
Bigger effects are expected at lower $Q^2$.

If QCD instantons contribute to the fireball event sample, then $\mathcal{T}$ would
be expected  to be different, possibly larger, for the fireball events than the typical 
value of about 0.07 or 0.08.
As this is not the case,
this measurement of $\mathcal{T}$ in photoproduction does not support the idea that QCD 
instantons contribute significantly to the production of the fireball events.
It should be noted that there is only a qualitative prediction on the contribution from QCD instantons
based on democracy of flavours, including heavy flavours, subject to available energy.  
The only existing calculation \cite{hep-ph-9906441} applies to DIS and only considers three massless flavours.  
Since there is no charm-quark contribution, this calculation is probably only valid at low particle multiplicities,
where the number of $K^{0}_{S}$ is predicted to be about twice as big as that predicted by {\sc Ariadne}. 
\subsection{Polarization}

In analogy with QED, the spin-orbit interaction leads to polarization in scattering due to the
strong interaction \cite{pl:b105:403,ijmp:a5:1197,pl:b183:357}.
Unpolarized $s$ quarks get partially transversely polarized due to elastic scattering in the coloured field along
the direction of {\bf $k_i \times k_f$}, where {\bf $k_i$} and {\bf $k_f$} stand for the initial
and final momenta of the $s$ quarks.
The degree of the polarization depends on the scattering angle and the strength of the coloured
field. In the constituent quark model, the $\Lambda$ inherits its spin from the $s$ quark, and 
{\bf $k_f$} is along the $\Lambda$ momentum. As {\bf $k_i$} is unknown in our case, the electron beam
direction was used instead (the effect of using the jet direction was also investigated). 

\begin{figure}[h]
\vfill
\begin{center}
\epsfig{file=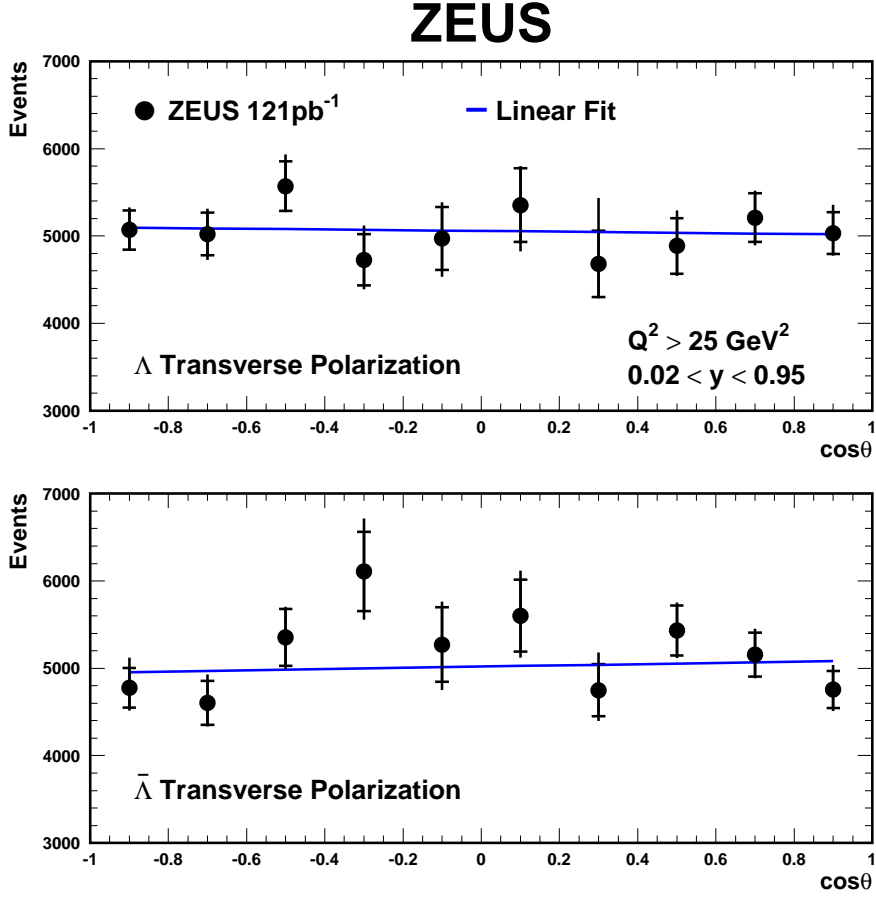,width=13cm}
\end{center}
\caption{
Angular distributions of the highest-momentum decay particle from a $\Lambda$ or $\bar{\Lambda}$ 
in the range $0.6 < P_{T}^\mathrm{LAB} < 2.5 \gev$ and $|\eta^\mathrm{LAB}| < 1.2$ for 
events with $Q^{2} > 25 \gev^2$ and $0.02 < y < 0.95$,
where $\theta$ is the angle between the decay-particle momentum vector and the polarization axis, in
the rest frame of the $\Lambda$ or $\bar{\Lambda}$.
Statistical errors (inner error bars) and the systematic uncertainties added in quadrature are shown.  
The first-order polynomial fit (solid line) from which the polarization is obtained is also shown.  
}
\label{tpol_hiq2}
\vfill
\end{figure}

The transverse polarization $\mathcal{P}^\Lambda$ ($\mathcal{P}^{\bar{\Lambda}}$) is defined 
by the form of the proton (antiproton) angular distribution:
\[
\frac{1}{N} \frac{\textrm{d} N}{\textrm{d} \cos\theta} = 
\frac{1}{2} [ 1 + \alpha \mathcal{P}^{\Lambda} \cos \theta],
\]
\[
\frac{1}{N} \frac{\textrm{d} N}{\textrm{d} \cos\theta} = 
\frac{1}{2} [ 1 - \alpha \mathcal{P}^{\bar{\Lambda}} \cos \theta],
\]
where  $\alpha$ is the decay asymmetry parameter, measured to be $\alpha = 0.642 \pm 0.013$ \cite{pl:b592:1}, 
and $\theta$ is the angle between the proton (antiproton) momentum boosted to the rest frame 
of the $\Lambda (\bar{\Lambda})$ and the polarization axis, 
$k_\mathrm{e}^\mathrm{beam} \times k_{\Lambda}$. 
An example of the angular distribution of the proton's (antiproton's) momenta with respect to 
the polarization axis,
boosted to the $\Lambda$($\bar{\Lambda}$) rest frame, is shown in Fig. \ref{tpol_hiq2}.

Fitted values of the transverse polarization $\mathcal{P}^\Lambda$ and 
$\mathcal{P}^{\bar{\Lambda}}$ are presented in Table \ref{tpol_values}
for high- and low-$Q^2$ DIS and for photoproduction. 
All values are compatible with no polarization.
Also presented are the polarization values obtained by investigating the 
angular distribution of the higher-momentum $\pi$ from $K^{0}_{S}$ decays, as a further test of any 
systematic detector effect. 

\section{Conclusions}

Measurements of $K^{0}_{S}$, $\Lambda$ and $\bar{\Lambda}$ production have been made at HERA, 
using 121$\pb^{-1}$ of data collected with the ZEUS detector. 
The following conclusions have been obtained:

\begin{enumerate}
\item in high- and low-$Q^2$ DIS, {\sc Ariadne} reproduces gross features of the cross sections 
but shows discrepancies in detail.  
Overall, the strangeness suppression factor $\lambda_s = 0.3$ is preferred to $\lambda_s = 0.22$.
{\sc Pythia}, normalised to the data, describes the dependence of the photoproduction cross sections on
$P_{T}^\mathrm{LAB}$ and $\eta^\mathrm{LAB}$ satisfactorily but fails to reproduce the $x_{\gamma}^\mathrm{OBS}$ 
dependence;
\item the numbers of $\Lambda$ and $\bar{\Lambda}$ produced are consistent with being equal;
\item except for the resolved photon interactions, the measured ratio of baryons to mesons
$\mathcal{R}$, defined as:
\begin{equation*} \label{baryon_ratio_eqn}
\mathcal{R} = \frac{N(\Lambda) + N(\bar{\Lambda})}{N(K^{0}_{S})},
\end{equation*}
is in the range between 0.2 and 0.5, similar to measurements at $e^+ e^-$ colliders. 
{\sc Ariadne} and {\sc Pythia} follow the shapes of the data on the selected observables 
but in many cases fail quantitatively at the 10 to 20\% level;
\item in the resolved photon region, the ratio of baryons to mesons is large, significantly larger than measured 
in $e^+ e^-$ interactions and significantly larger than the {\sc Pythia} prediction; 
\item the ratio of strange-to-light hadrons measured 
in terms of $\mathcal{T}$:
\begin{equation*} \label{to track_ratio_eqn}
\mathcal{T} = \frac{N(K^{0}_{S})}{N_\mathrm{ch}},
\end{equation*}
is compatible with measurements at other colliders 
and is described by {\sc Ariadne} and {\sc Pythia} for all investigated samples of events. For the DIS 
sample, the strangeness suppression factor $\lambda_s = 0.22$ is preferred to the default value of 
$\lambda_s = 0.3$. For the photoproduction sample, {\sc Pythia} with $\lambda_s = 0.3$ overestimates the data, 
but describes the shape of the distributions. There is no indication of any unusual yield of strange-hadrons 
in the fireball-enriched sample, as would be qualitatively expected had there been a significant contribution 
from QCD instantons;
\item No evidence has been found for non-zero transverse polarization in inclusive 
$\Lambda$ or $\bar{\Lambda}$ production.
\end{enumerate}

\section{Acknowledgments}
We thank the DESY directorate for its strong support and encouragement.
The remarkable achievements of the HERA machine group were essential for the successful completion
of this work and are greatly appreciated.  
The design, construction and installation of the ZEUS detector have been made possible by the efforts 
of many people who are not listed as authors.  
%

\providecommand{\etal}{et al.\xspace}
\providecommand{\coll}{Coll.\xspace}
\catcode`\@=11
\def\@bibitem#1{%
\ifmc@bstsupport
  \mc@iftail{#1}%
    {;\newline\ignorespaces}%
    {\ifmc@first\else.\fi\orig@bibitem{#1}}
  \mc@firstfalse
\else
  \mc@iftail{#1}%
    {\ignorespaces}%
    {\orig@bibitem{#1}}%
\fi}%
\catcode`\@=12

%

\begin{table}[!htp]
\centering
\hspace*{-15mm}
\begin{center}
\begin{tabular}{|c|c|c|c|}
\hline
\rule[-3mm]{0mm}{8mm} &  \multicolumn{3}{c|}{Polarization (\%)} \\
\hline
 \rule[-3mm]{0mm}{8mm} &  High-$Q^{2}$ DIS & Low-$Q^{2}$ DIS &  Photoproduction \\
\hline
\rule[-3mm]{0mm}{8mm} $\Lambda$ & $-1.3 \pm 4.3(stat.)^{+4.0}_{-0.8}(syst.)$ & $-4.0 \pm 5.3(stat.)^{+4.7}_{-4.0}(syst.)$ & $-2.4 \pm 2.2(stat.)$  \\
\hline
\rule[-3mm]{0mm}{8mm} $\bar{\Lambda}$ & $-2.2 \pm 4.2(stat.)^{+2.4}_{-1.3}(syst.)$ & $-8.5 \pm 5.5(stat.)^{+4.7}_{-2.1}(syst.)$ & $-5.8 \pm 2.2(stat.)$ \\
\hline
\rule[-3mm]{0mm}{8mm} $K^{0}_{S}$ & $-1.5 \pm 1.1(stat.)$ & $-0.05 \pm 1.5(stat.)$ & $-0.5 \pm 0.2(stat.)$ \\
\hline
\end{tabular}
\caption{The transverse polarization values for $\Lambda$ and $\bar{\Lambda}$, expressed here in \%, in the 
high-$Q^{2}$ DIS ( $Q^{2} > 25 \gev^2$ and $0.02 < y < 0.95$), 
low-$Q^{2}$ DIS ( $5 < Q^{2} < 25 \gev^2$ and $0.02 < y < 0.95$), 
and photoproduction ( $Q^{2} < 1 \gev^2$, $0.2 < y < 0.85$ and with two jets $E_{T}^{\mathrm{jet}} > 5 \gev$ 
and $|\eta^{\mathrm{jet}}|<2.4$) samples. 
Only $\Lambda$ and $\bar{\Lambda}$ in the range $0.6 < P_{T}^\mathrm{LAB} < 2.5 \gev$ and $|\eta^\mathrm{LAB}| < 1.2$ are considered.
The statistical error is quoted for all samples, together with the systematic uncertainty associated with the 
measurement for the high-$Q^{2}$ and low-$Q^{2}$ samples.  
A similar systematic uncertainty is expected for the photoproduction sample.  
Also shown, as a test of any systematic effect, are the polarization values obtained by investigating the 
angular distribution of the higher-momentum $\pi$ from $K^{0}_{S}$ decays. 
\label{tpol_values}}
\end{center}
\end{table}

\newpage

\newpage

\newpage

\newpage

\newpage

\newpage
\clearpage

\newpage

\newpage

\newpage

\newpage

\newpage

\newpage

\newpage

\newpage

\clearpage

\end{document}